\documentclass[english,tightenlines,eqsecnum,amsmath,amssymb,aps,prd,superscriptaddress,nofootinbib,showpacs,floats]{article}
\usepackage{amsmath,amsfonts,amssymb,multicol}
\usepackage{graphicx,caption,subcaption}
\usepackage{epstopdf}
\usepackage[usenames]{xcolor}
\usepackage{epstopdf}
\definecolor{AHZ}{rgb}{0.0,0.9,0.2}
\usepackage[colorlinks,linkcolor=AHZ,citecolor=red,bookmarks]{hyperref}
\usepackage{rotating}
\usepackage[mathscr]{eucal}
\usepackage{graphicx}
\usepackage[T1]{fontenc}
\usepackage{babel}
\usepackage{authblk}
\usepackage{epsfig}
\usepackage{color}
\usepackage{amssymb}
\usepackage{fancyhdr}

\def\nn{\nonumber\\}
\newcommand{\f}[2]{\frac{#1}{#2}}
\def\be{\begin{equation}}
\def\ee{\end{equation}}
\def\bea{\begin{eqnarray}}
\def\eea{\end{eqnarray}}

\def\bwt{\begin{widetext}}
\def\ewt{\end{widetext}}

\usepackage{amsmath}

\usepackage{amssymb}

\begin{document}

\title{Stability of the Einstein static Universe in Einstein-Cartan-Brans-Dicke gravity}
\author[1,2]{Hamid Shabani\thanks{H.Shabani@phys.usb.ac.ir}}
\author[2]{Amir Hadi Ziaie\thanks{ah.ziaie@riaam.ac.ir}}
\affil[1]{Physics Department, Faculty of Sciences, University of Sistan and Baluchestan, Zahedan, Iran}
\affil[2]{Research~Institute~for~Astronomy~and~Astrophysics~of~ Maragha~(RIAAM), P.O.~Box~55134-441,~Maragha,~Iran}
\renewcommand\Authands{ and }
\maketitle
\begin{abstract}
In the present work we consider the existence and stability of Einstein static ({\sf ES}) Universe in Brans-Dicke ({\sf BD}) theory with non-vanishing spacetime torsion. In this theory, torsion field can be generated by the {\sf BD} scalar field as well as the intrinsic angular momentum (spin) of matter. Assuming the matter content of the Universe to be a  Weyssenhoff fluid, which is a generalization of perfect fluid in general relativity ({\sf GR}) in order to include the spin effects, we find that there exists a stable {\sf ES} state for a suitable choice of the model parameters. We analyze the stability of the solution by considering linear homogeneous perturbations and discuss the conditions under which the solution can be stable against these type of perturbations. Moreover, using dynamical system techniques and numerical analysis, the stability regions of the {\sf ES} Universe are parametrized by the {\sf BD} coupling parameter and first and second derivatives of the {\sf BD} scalar field potential, and it is explicitly shown that a large class of stable solutions exists within the respective parameter space. This allows for non-singular emergent cosmological scenarios where the Universe oscillates indefinitely about an initial {\sf ES} solution and is thus past eternal. 
\end{abstract}
\section{Introduction}
It is well known that inflation, a short-lived and prompt accelerated cosmic expansion era in the very early Universe, is a successful theory in solving some problems from which the standard hot big bang cosmology suffers. Inflationary cosmology predicts a nearly scale-invariant power spectrum for primordial curvature perturbations, which was confirmed by cosmic microwave background ({\sf CMB}) observations \cite{CMB}. In spite of its successes, the inflationary scenario still suffers from the problem of initial singularity of the Universe. Some models have been proposed so far in order to cure this problem, i.e., some mechanisms within
the framework of quantum gravity such as pre-big bang~\cite{prebigb} and cyclic scenarios \cite{cyclicuni} in string/M theory. Moreover, it has been shown that bouncing cosmology~\cite{Bouncecos} and emergent Universe ({\sf EU}) scenario~\cite{Ellis-Maartens2004,MinSFESU}, as an alternative to cosmic inflation, can also avoid the big bang singularity. The concept of {\sf EU} is a very interesting idea in standard cosmology with the aim of searching for singularity free inflationary models. In this model, our Universe has no time-like singularity, it is ever existing and has almost a static behavior (Einstein static ({\sf ES}) state) in the infinite past ($t\rightarrow-\infty$)  and then evolves into an inflationary era. The Universe is then originated from the {\sf ES} state rather than the initial big bang singularity. During the last decades, many authors have proposed {\sf ES} Universe within different scenarios as it provides a setting in which the Universe is ever existing and large enough so that the spacetime may be treated at classical levels. In 1930, Eddington studied the stability of {\sf ES} solutions in general relativity ({\sf GR}) and found that these solutions are unstable against homogeneous and isotropic perturbations~\cite{EddES1930}. In 1967, Harrison introduced a model of closed Universe stuffed from radiation in the presence of a cosmological constant, which asymptotically coincides with the {\sf ES} model in the infinite past~\cite{Harrison1967} but the scenario does not exit to an inflationary phase. Later studies carried out by Gibbons~\cite{Gibbons1987} and Barrow and his coworkers~\cite{Barrowetal2003} found that the {\sf ES} Universe
with a perfect fluid is neutrally stable with respect to small inhomogeneous vector and tensor linear perturbations, and against scalar perturbations if the sound speed of perfect fluid fulfills $c_s^2>1/5$. A similar cosmological scenario was considered by Ellis and Maartens where the possibility of avoiding the big-bang singularity without resorting to a quantum regime for spacetime, was investigated~\cite{Ellis-Maartens2004}. Work along this line has been performed within different models among which we can quote, a closed Universe with a minimally coupled scalar field $\phi$ and self-interacting potential $V(\phi)$~\cite{MinSFESU}, {\sf EU} filled with exotic matter~\cite{Mukherjee2006}, brane-world scenario~\cite{braneemergent}, Einstein-Gauss-Bonnet theory~\cite{Gaussemergent}, $f({\sf R})$ theory~\cite{FRemergent}, $f({\sf T})$ gravity~\cite{FTemergent}, loop quantum cosmology~\cite{lqcemergent} and other gravity theories~\cite{otheremergent}. Moreover, the study of this subject in non-Riemannian spacetimes has been done and recently, the existence and stability of an {\sf ES} Universe in the framework of Einstein-Cartan ({\sf EC}) theory has been investigated in~\cite{KhedmatJCAP}. The author has shown that for a spatially closed Universe filled with a Weyssenhoff spinning fluid there is a stable {\sf ES} state and the Universe can live at this state past-eternally. However, as this stable state corresponds to a center equilibrium point, the Universe may not naturally evolve from it into an inflationary phase. Further studies and developments on this issue has been done in~\cite{ECEmergent} where it is shown that an emergent scenario that avoids the initial singularity of the Universe can be successfully implemented in {\sf EC} theory.
\par
The {\sf ES} model has been also studied in the well-known Brans-Dicke ({\sf BD}) theory and it is shown that a stable past-eternal static solution can be obtained that eventually enters into an unstable phase where the stability of the solution is broken leading to an inflationary period~\cite{emergentBDT}. The {\sf BD} theory which is an alternative gravity theory is a natural generalization of {\sf GR} where the gravitational coupling constant is replaced by a scalar field~\cite{BDTpp}. This theory in which, gravitational interactions are described by the metric of a Riemannian spacetime and a non-minimally coupled scalar field on that spacetime, is an attempt towards improving {\sf GR} from the standpoint of Mach's principle~\cite{BDTpp,BDTpp1}. The {\sf BD} theory has received much interests as it appears naturally in the models dealing with supergravity, Kaluza-Klein theories and in the low energy limit of string theories~\cite{BDTpp2}. Recently, the modified {\sf BD} theory has been investigated on a general spacetime manifold with non-vanishing torsion field~\cite{BDTTOR1986} and it was shown that both the scalar field and spin of fermionic particles have contribution in generating the spacetime torsion field. This theory can be viewed as a sort of unification of BD theory and {\sf EC} theory, i.e., the simplest Poincare gauge theory of gravity, in the framework of which, the gravitational interactions are described by means of spacetime curvature and torsion with the sources being energy-momentum and spin tensors~\cite{PGTHehl2013}. It is well known that the field equation for spacetime torsion in {\sf EC} theory is purely algebraically coupled to the spin tensor of matter and therefore the presence of torsion is restricted by spinning matter distribution, thus the spacetime torsion cannot propagate outside the matter through the vacuum~\cite{Venzo-1-Hehl,Venzo-2-Hehl}. However, in {\sf BD} theory with torsion (that hereafter we call it as Einstein-Cartan-Brans-Dicke ({\sf ECBD}) theory) there exists an interesting possibility where a varying gravitational coupling could acts as a source of spacetime torsion. Thus from the standpoint of {\sf ECBD} theory, the {\sf BD} scalar field can play the role of a mediator field i.e., from one side, it participates within the gravitational interactions through its non-minimal coupling to curvature and from another side, it behaves as a source, alongside with the spin of fermionic matter, for spacetime torsion field. During the last years, some authors have studied cosmological as well as astrophysical aspects of {\sf ECBD} theory such as, static spherically symmetric spacetimes in vacuum where it is shown that torsion field can propagate via the scalar field even if the spin angular momentum is absent~\cite{Sung-Won KIM1987}. Cosmological models in the framework of {\sf ECBD} theory has been built and investigated in~\cite{ECBDCOSMOD} and higher dimensional extension of {\sf ECBD} theory has been studied in~\cite{HIGHDECBD} where it is shown that the electromagnetic field and the scalar field appear during the reduction of five-dimensional action. In the case of extreme phenomena where the regimes of ultra-strong gravity are present e.g., the late stages of the gravitational collapse scenario, the BD scalar field may have some effects on stellar structure and final fate of the collapse process~\cite{BDCOLLAPSE}. Moreover, from the standpoint of cosmological implications, it has been shown that the {\sf BD} scalar field, though may be undetectable at the present epoch, could play an important role in the very early Universe~\cite{BDVEU} (see also \cite{SINGHRAI1983} and references therein). It is therefore of interest to study the possibility of existence and stability of {\sf ES} Universe in {\sf ECBD} theory, as the very early Universe was a hot soup of the fundamental particles including fermionic fields and thus torsion field could be present due to the spin effects of fermions~\cite{Venzo-1-Hehl,Venzo-2-Hehl,SPINTORSIONREF,SPINTORSIONREF1}. In the present work, Motivated by the these considerations, we seek for the stable solutions representing an {\sf ES} state for the Universe in the framework of {\sf ECBD} theory. Our paper is then organized as follows: In section~\ref{ECNDFEQSder} we derive the field equations of {\sf ECBD} theory considering a Weyssenhoff fluid for the matter content of the Universe. In section~\ref{stabecbd}, we give the conditions for the stable {\sf ES} solution and proceed with analyzing this solution using dynamical system approach in section~\ref{dynsysappes}. In section~\ref{summconc}, we summarize and discuss our results.

\section{Field equations of {\sf ECBD} theory}\label{ECNDFEQSder}
Let us assume a Riemann-Cartan manifold as the background spacetime which is endowed with the {\sf BD} scalar field $\Phi$. The spacetime torsion is defined as the antisymmetric part of the general connection $\tilde{\Gamma}^{\alpha}_{\,\,\beta\gamma}$ given by
\bea\label{tordef}
{\sf Q}^{\alpha}_{\,\,\,\,\beta\mu}=\tilde{\Gamma}^{\alpha}_{~\beta\mu}-\tilde{\Gamma}^{\alpha}_{~\mu\beta}.
\eea
We assume that the metricity condition $\tilde{\nabla}_{\alpha}g_{\mu\nu}=0$ holds in {\sf ECBD} which leads to the following relation between the general connection ($\tilde{\Gamma}^{\alpha}_{\,\,\beta\gamma}$) and the Christoffel connection ($\Gamma^{\alpha}_{\,\,\beta\gamma}$) as
\be\label{GAFC}
\tilde{\Gamma}^{\alpha}\!\!~_{\beta\gamma}=\Gamma^{\alpha}_{\,\,\beta\gamma}+{\sf K}^{\alpha}\!\!~_{\beta\gamma},
\ee
where the contorsion tensor is defined as 
\be\label{contortion}
{\sf K}^{\mu}_{~\alpha\beta}=\f{1}{2}\left({\sf Q}^{\mu}_{~\alpha\beta}-{\sf Q}_{\alpha~\beta}^{~\mu}-
{\sf Q}_{\beta~\alpha}^{~\mu}\right).
\ee
The action in {\sf ECBD} theory is written as 
\bea\label{action}
S&=&\int \sqrt{-{\sf g}} d^{4} x \left[\f{\Phi \tilde{{\sf R}}}{\kappa^2}-\f{\omega}{\Phi}{\sf g}_{\alpha\beta}\tilde{\nabla}^\alpha\Phi\tilde{\nabla}^\beta\Phi-V(\Phi)+{\sf L}_m\left({\sf g}_{\alpha\beta},{\sf K}^\alpha_{\,\,\,\,\beta\gamma},\Psi^i\right)\right],\nn
&=&\int \sqrt{-{\sf g}} d^{4}x\bigg\{\f{1}{\kappa^2}\bigg[\Phi {\sf R}(\Gamma)+\left({\sf K}^{\rho\lambda}_{\,\,\,\,\,\lambda}-{\sf K}^{\lambda\rho}_{\,\,\,\,\,\lambda}\right)\nabla_{\rho}\Phi+\Phi\left({\sf K}^{\sigma\rho}_{\,\,\,\,\,\nu}{\sf K}^{\nu}_{\,\,\,\sigma\rho}-{\sf K}^{\sigma\mu}_{\,\,\,\,\,\mu}{\sf K}^{\lambda}_{\,\,\,\sigma\lambda}\right)\bigg]
\nn&-&\f{\omega}{\Phi}{\sf g}_{\alpha\beta}{\nabla}^\alpha\Phi{\nabla}^\beta\Phi-V(\Phi)+{\sf L}_m\bigg\},
\eea
where $\kappa^2=\f{16\pi}{c^4}$ and ${\sf L}_m$ being the gravitational coupling constant and Lagrangian of minimally coupled matter field(s) $\Psi$, which generally depends on metric, spacetime torsion and their derivatives~\cite{Venzo-2-Hehl}. The symbol $\tilde{{\sf R}}$ being the Ricci curvature scalar constructed out of the general connection $\tilde{\Gamma}^{\alpha}_{\,\,\beta\gamma}$ and is given by
\bea\label{RICCIF}
\tilde{{\sf R}}={\sf R}(\Gamma)+\nabla_{\lambda}{\sf K}^{\lambda\rho}\!\!~_{\rho}-\nabla_{\rho}{\sf K}^{\lambda\rho}\!\!~_{\lambda}+{\sf K}^{\sigma\mu}\!\!~_{\mu}{\sf K}^{\lambda}\!\!~_{\sigma\lambda}
-{\sf K}^{\sigma\rho}\!\!~_{\nu}{\sf K}^{\nu}\!\!~_{\sigma\rho},
\eea
where $\tilde{\nabla}_\alpha$ and $\nabla_\alpha$ denote covariant derivatives with respect to $\tilde{\Gamma}^{\alpha}_{\,\,\beta\gamma}$ and $\Gamma^{\alpha}_{\,\,\beta\gamma}$ respectively. We note that in the first line of the action (\ref{action}), the covariant derivative on the {\sf BD} scalar field could be rewritten in terms of the usual covariant derivative and in the second line, we have used by part integration and have omitted total derivative terms in order to transfer the covariant derivative from contorsion terms to the {\sf BD} scalar field. The field equations can be obtained by independent variation of the action (\ref{action}) with respect to three independent fields, i.e., the metric and the spacetime torsion fields and the {\sf BD} scalar field. Varying action with respect to the  metric field gives
\bea\label{fieldeqsmetric}
&-&\!\!\!\!\Phi\,{\sf G}_{\alpha\beta}-\f{\Phi}{2}\bigg[{\sf K}^{\gamma\,\,\delta}_{\,\,\beta}\,{\sf K}_{\delta\gamma\alpha}+{\sf K}^{\gamma\,\,\delta}_{\,\,\alpha}\,{\sf K}_{\delta\gamma\beta}-{\sf g}_{\alpha\beta}{\sf K}^{\gamma\delta\epsilon}\,{\sf K}_{\delta\epsilon\gamma}+{\sf g}_{\alpha\beta}{\sf K}^{\gamma\delta}_{\,\,\,\,\,\gamma}\,{\sf K}_{\delta\,\,\epsilon}^{\,\,\epsilon}-{\sf K}^{\gamma}_{\,\,\alpha\beta}\,{\sf K}^{\delta}_{\,\,\gamma\delta}\nn&-&\!\!\!\!{\sf K}^{\gamma}_{\,\,\beta\alpha}\,{\sf K}^{\delta}_{\,\,\gamma\delta}\bigg]+\f{1}{2}{\sf K}^{\gamma}_{\,\,\beta\gamma}\nabla_\alpha\Phi+\f{1}{2}{\sf K}^{\gamma}_{\,\,\alpha\gamma}\nabla_\beta\Phi+\nabla_\alpha\nabla_\beta\Phi+\f{\omega}{\Phi}\nabla_\alpha\Phi\nabla_\beta\Phi\nn&+&\!\!\!\!\f{1}{2}\left[{\sf K}^{\gamma}_{\,\,\alpha\beta}-{\sf K}^{\gamma}_{\,\,\beta\alpha}\right]\nabla_\gamma\Phi+\f{1}{2}{\sf g}_{\alpha\beta}\left[{\sf K}^{\delta\gamma}_{\,\,\,\,\gamma}-{\sf K}^{\gamma\delta}_{\,\,\,\,\gamma}\right]\nabla_\delta\Phi+{\sf g}_{\alpha\beta}\Box\Phi\nn&-&\!\!\!\!\f{\omega}{2\Phi}{\sf g}_{\alpha\beta}\nabla_\gamma\Phi\nabla^\gamma\Phi-\f{1}{2}{\sf g}_{\alpha\beta}V(\Phi)=\f{\kappa^2}{2}{\sf T}_{\alpha\beta},
\eea
where ${\sf T}_{\alpha\beta}=2\left(\delta\sf {L}_m/\delta {\sf g}^{\alpha\beta}\right)/\sqrt{-{\sf g}}$ being the energy-momentum tensor ({\sf EMT}) of matter fields. Varying the action with respect to contorsion gives the modified Cartan field equation as
\bea\label{modcarteq}
\Phi\left[{\sf K}_{\beta\gamma\alpha}+{\sf K}_{\gamma\alpha\beta}-g_{\alpha\gamma}{\sf K}_{\beta\,\,\,\delta}^{\,\,\,\delta}-g_{\beta\gamma}{\sf K}^{\delta}_{\,\,\alpha\delta}\right]+{\sf g}_{\beta\gamma}\nabla_\alpha\Phi-{\sf g}_{\alpha\gamma}\nabla_\beta\Phi=\f{\kappa^2}{2}\tau_{\alpha\beta\gamma},
\eea
where $\tau^{\alpha\beta\gamma}=2\left(\delta(\sqrt{-g}{\sf L}_m)/\delta {\sf K}_{\alpha\beta\gamma}\right)/\sqrt{-g}$ is defined as the spin tensor of matter~\cite{Venzo-1-Hehl}. Next, we use the expression (\ref{contortion}) to rewrite the above equation as
\bea\label{modcartrew}
\Phi\left[{\sf Q}_{\gamma\alpha\beta}+{\sf g}_{\alpha\gamma}{\sf Q}_\beta-{\sf g}_{\beta\gamma}{\sf Q}_\alpha\right]+{\sf g}_{\beta\gamma}\nabla_\alpha\Phi-{\sf g}_{\alpha\gamma}\nabla_\beta\Phi=\kappa^2\tau_{\alpha\beta\gamma},
\eea
where ${\sf Q}_\mu={\sf Q}^{\alpha}_{\,\,\,\mu\alpha}$ being the trace of torsion tensor. Let us now proceed with employing a classical description of spin as postulated by Weyssenhoff, which is given by \cite{KCQG1987W1947}
\be\label{WeyFC}
\tau_{\gamma\beta\alpha}={\sf S}_{\beta\alpha}{\sf U}_{\gamma},~~~~~~~~{\sf S}_{\mu\nu}{\sf U}^{\mu}=0,
\ee
where ${\sf U}^{\alpha}$ is the four-velocity of the fluid element and ${\sf S}_{\mu\nu}=-{\sf S}_{\nu\mu}$ is a second-rank antisymmetric tensor which is defined as the spin density tensor. Its spatial components include the 3-vector $({\sf S}^{23},{\sf S}^{13},{\sf S}^{12})$ which coincides in the rest frame with the spatial spin density of the matter element. The left spacetime components
$({\sf S}^{01}, {\sf S}^{02}, {\sf S}^{03})$ are assumed to be zero in the rest frame of fluid element, which can be covariantly formulated as the constraint given in the second part of (\ref{WeyFC}). This constraint on the spin density tensor is usually called the Frenkel condition which requires the intrinsic spin of matter to be spacelike in the rest frame of the fluid. Contracting equation (\ref{modcartrew}) with ${\sf g}^{\gamma\beta}$ gives
\bea\label{Torphirel}
{\sf Q}_{\alpha}=\f{3}{2}\f{\nabla_\alpha\Phi}{\Phi},
\eea
where use has been made of the second expression in (\ref{WeyFC}). Substituting then for the trace of torsion into equation (\ref{modcartrew}) we finally get the desired relations for the torsion and contorsion tensors as
\bea\label{finaleqtor}
{\sf Q}^{\gamma}_{\,\,\,\alpha\beta}=\f{\kappa^2}{2\Phi}{\sf U}^{\gamma}{\sf S}_{\beta\alpha}+\f{1}{2\Phi}\left[\delta_{\beta}^{\,\,\gamma}\nabla_\alpha\Phi-\delta_\alpha^{\,\,\gamma}\nabla_\beta\Phi\right],
\eea
and
\bea\label{finaleqcontor}
{\sf K}^{\mu}_{\,\,\,\alpha\beta}=-\f{\kappa^2}{4\Phi}\left[{\sf S}_\beta^{\,\,\,\mu}{\sf U}_\alpha+{\sf S}_\alpha^{\,\,\mu}{\sf U}_\beta+{\sf S}_{\alpha\beta}{\sf U}^\mu\right]+\f{1}{2\Phi}\left[\delta_\beta^{\,\,\mu}\nabla_\alpha\Phi-{\sf g}_{\alpha\beta}\nabla^{\mu}\Phi\right].
\eea
It is noteworthy that in {\sf ECBD} theory, the spin distribution of fermions is not the only source of spacetime torsion and the {\sf BD} scalar field also contributes to generating torsion field. Varying action (\ref{action}) with respect to $\Phi$ gives the evolution equation for the {\sf BD} scalar field as
\bea\label{evoleqbdfiled1}
&&{\sf K}^{\alpha\beta\gamma}\,{\sf K}_{\beta\gamma\alpha}-{\sf K}^{\alpha\beta}_{\,\,\,\,\,\alpha}\,{\sf K}_{\beta\,\,\,\gamma}^{\,\,\,\gamma}+{\sf R}-\nabla_\alpha{\sf K}^{\alpha\beta}_{\,\,\,\,\,\beta}+\nabla_\beta{\sf K}^{\alpha\beta}_{\,\,\,\,\,\alpha}\nn&&+\f{2\omega}{\Phi}\Box\Phi-\f{\omega}{\Phi^2}\nabla_\alpha\Phi\nabla^\alpha\Phi-\f{dV(\Phi)}{d\Phi}=0.
\eea
The stress-energy tensor at the right hand side of equation (\ref{fieldeqsmetric}) can be decomposed into the usual perfect fluid part, ${\sf T}^{\sf PF}_{\alpha\beta}$, and an intrinsic spin part, ${\sf T}^{\sf SF}_{\alpha\beta}$ as \cite{gasprl}
\bea\label{stentenpfsf}
{\sf T}_{\alpha\beta}&=&{\sf T}^{\sf PF}_{\alpha\beta}+{\sf T}^{\sf SF}_{\alpha\beta}= \left[(\rho + p){\sf U}_{\alpha}{\sf U}_{\beta}-p{\sf g}_{\alpha\beta}\right]+\f{1}{2}{\sf Q}_{\nu\mu(\alpha}{\sf S}^{\mu}_{\,\,\,\beta)}{\sf U}^\nu\nn&+&{\sf U}_{(\alpha}{\sf S}_{\beta)\mu}\,{\sf K}^{\rho\mu}_{\,\,\,\,\,\,\nu}\,{\sf U}_\rho+{\sf U}^\rho{\sf K}^\mu_{\,\,\,\sigma\rho}{\sf U}^\sigma{\sf U}_{(\alpha}{\sf S}_{\beta)\mu}-\f{1}{2}{\sf U}_{(\alpha}{\sf Q}_{\beta)\mu\nu}{\sf S}^{\mu\nu},
\eea
whence, using expressions (\ref{finaleqtor}) and (\ref{finaleqcontor}) we get
\bea\label{spflutensor}
{\sf T}_{\alpha\beta}&=&\f{\kappa^2}{8\Phi}\left[2{\sf S}_{\gamma\delta}\,{\sf S}^{\gamma\delta}{\sf U}_{\alpha}{\sf U}_{\beta}-2{\sf S}_{\alpha}^{\,\,\,\delta}{\sf S}_{\beta\delta}{\sf U}_\gamma{\sf U}^{\gamma}\right]+\f{\nabla^\gamma\Phi}{8\Phi}\bigg[{\sf S}_{\beta\gamma}{\sf U}_\alpha+{\sf S}_{\alpha\gamma}{\sf U}_\beta\bigg]\nn&+& \left[(\rho + p){\sf U}_{\alpha}{\sf U}_{\beta}-p{\sf g}_{\alpha\beta}\right],
\eea
where use has been made of the Frenkel condition. In order to rewrite the field equation (\ref{fieldeqsmetric}) we can substitute from equation (\ref{finaleqcontor}) for the contorsion tensor into this equation which gives
\bea\label{fieldeqsmetricrewrite}
&-&\Phi{\sf G}_{\alpha\beta}+\f{\kappa^4}{16\Phi}\left[{\sf S}_{\gamma\delta}{\sf S}^{\gamma\delta}{\sf U}_\alpha{\sf U}_\beta-\f{1}{2}{\sf g}_{\alpha\beta}{\sf S}_{\delta\epsilon}{\sf S}^{\delta\epsilon}{\sf U}_\gamma{\sf U}^\gamma\right]+\nabla_{\alpha}\nabla_\beta\Phi\nn&+&\f{2\omega+3}{2\Phi}\left[\nabla_{\alpha}\Phi\nabla_{\beta}\Phi-\f{1}{2}{\sf g}_{\alpha\beta}\nabla_\gamma\Phi\nabla^\gamma\Phi\right]+\f{\kappa^2}{8\Phi}\bigg[{\sf S}_{\beta\gamma}\,{\sf U}_\alpha\,\nabla^\gamma\Phi+{\sf S}_{\alpha\gamma}\,{\sf U}_\beta\,\nabla^\gamma\Phi\bigg]\nn&-&{\sf g}_{\alpha\beta}\Box\Phi-\f{1}{2}{\sf g}_{\alpha\beta}V(\Phi)={\sf T}_{\alpha\beta}.
\eea
From the microscopical point of view, a randomly oriented gas of fermionic particles is the source for the spacetime torsion. However, the effective sources for the macroscopic gravitational field are to be treated at macroscopic level, thus, a suitable space-time averaging on {\sf EMT} and spin sources in (\ref{fieldeqsmetricrewrite}) has to be carried out~\cite{Venzo-1-Hehl}. In this regard, if the spin orientation of particles is random,average of the spin density tensor and its derivative vanish, $\langle {\sf S}_{\mu\nu} \rangle=0$ and $\langle \nabla_{\alpha}{\sf S}_{\mu\nu} \rangle=0$~\cite{SPINTORSIONREF1,gasprl,poplawski2010}. Despite the vanishing of this term macroscopically, the square of spin density tensor, as appeared within the  square brackets of the first term at the right hand side of expression (\ref{spflutensor}) would have contribution to the field equations. Thus, for an unpolarized fermionic gas,
the averaging procedure gives~\cite{gasprl}
\bea\label{avgprospinsq}
\langle {\sf S}_{\mu\nu}{\sf S}^{\mu\nu}\rangle=2{\sf \sigma}^2,\nn
\langle{\sf S}_\mu^{\,\,\beta}{\sf S}_{\nu\beta}\rangle=\f{2}{3}\left[{\sf g}_{\mu\nu}-{\sf U}_\mu{\sf U}_\nu\right]{\sf \sigma}^2.
\eea
From equations (\ref{spflutensor}) and (\ref{fieldeqsmetricrewrite}) along with the above expressions for the square of spin density tensor, we finally get the modified {\sf BD} field equation in the presence of spin effects as 
\bea
{\sf G}_{\alpha\beta}&=&\f{\kappa^2}{\Phi}\left[\left(\rho+p-\f{\kappa^2\sigma^2}{6\Phi}\right){\sf U}_{\alpha}{\sf U}_{\beta}-{\sf g}_{\alpha\beta}\left(p-\f{\kappa^2\sigma^2}{12\Phi}\right)\right]+\f{3+2\omega}{2\Phi^2}\nabla_{\alpha}\Phi\nabla_{\beta}\Phi\nn&+&\f{1}{\Phi}\nabla_{\alpha}\nabla_{\beta}\Phi-{\sf g}_{\alpha\beta}\left[\f{\Box\Phi}{\Phi}+\f{3+2\omega}{4\Phi^2}\nabla_{\epsilon}\Phi\nabla^{\epsilon}\Phi+\f{V(\Phi)}{2\Phi}\right].\label{gmunu}
\eea
Next, we proceed with finding the evolution equation for {\sf BD} scalar field. To this aim we contract the above equation with metric tensor in order to find the Ricci scalar and then substitute for it into equation (\ref{evoleqbdfiled1}) which gives
\bea
\Box\Phi=\f{\kappa^2}{2(3+\omega)}\left[\rho-3p-\f{3\kappa^2\sigma^2}{8\Phi}\right]+\f{1}{6+2\omega}\left(\Phi\f{dV}{d\Phi}-2V(\Phi)\right).\label{eqelolsf}
\eea
In order to find an {\sf ES} solutions we consider a homogeneous and isotropic Universe described by a spatially non-flat Friedmann-Lema\^{\i}tre-Robertson-Walker ({\sf FLRW}) spacetime whose line element can be parametrized as
\begin{align}\label{metricFRW}
ds^{2}=dt^{2}-a^{2}(t) \Big{(}\frac{dr^{2}}{1-kr^2}+r^{2}d\Omega^2\Big{)}.
\end{align}
The field equations can then be written as (we use the units so that $\kappa^2=1$~\cite{SINGHRAI1983})
\bea\label{ecbdfes1}
3\left[H^2+\f{k}{a^2}\right]=\f{1}{\Phi}\left[\rho-\f{\sigma^2}{12\Phi}\right]+\f{2\omega+3}{4}\f{\dot{\Phi}^2}{\Phi^2}-3H\f{\dot{\Phi}}{\Phi}-\f{V(\Phi)}{2\Phi},~~~H=\f{d{a}/dt}{a}=\f{\dot{a}}{a},
\eea
\bea\label{ecbdfes2}
2\f{\ddot{a}}{a}+\f{\dot{a}^2}{a^2}+\f{k}{a^2}=-\f{1}{\Phi}\left[p-\f{\sigma^2}{12\Phi}\right]-2H\f{\dot{\Phi}}{\Phi}-\f{2\omega+3}{4}\f{\dot{\Phi}^2}{\Phi^2}-\f{\ddot{\Phi}}{\Phi}-\f{V(\Phi)}{2\Phi},
\eea
\bea\label{evoleqsf0}
\ddot{\Phi}+3H\dot{\Phi}=\f{1}{2\omega+6}\left[\rho-3p-\f{3\sigma^2}{8\Phi}\right]+\f{1}{2\omega+6}\left[\Phi\f{dV}{d\Phi}-2V\right].
\eea
We also have the following conservation equations for fluid part and spin part as
\bea
\f{d\rho}{dt}+3H(\rho+p)=0,\label{conseeqs1}\\
\f{d\sigma^2}{dt}+6H\sigma^2=0\label{conseeqs2}.
\eea

\section{Static Universe in {\sf ECBD}}\label{stabecbd}
The static solution in {\sf ECBD} model is a closed Universe characterized by the conditions $a=a_{\sf ES}={\rm constant}$, $\dot{a}_{\sf ES}=\ddot{a}_{\sf ES}=0$ for the scale factor and its derivatives, $\Phi=\Phi_{\sf ES}={\rm constant}$ and $\dot{\Phi}_{\sf ES}=\ddot{\Phi}_{\sf ES}=0$ for the {\sf BD} scalar field. Then, from equations (\ref{ecbdfes1})-(\ref{evoleqsf0}) for a spatially closed {\sf FLRW} Universe ($k=1$) and taking the equation of state ({\sf EoS}) as $p=(\gamma-1)\rho$, we get 
\bea
\f{3}{a_{\sf ES}^2}\!\!\!&=&\!\!\!\f{1}{\Phi_{\sf ES}}\left[\rho_{\sf ES}-\f{\sigma_{\sf ES}^2}{12\Phi_{\sf ES}}\right]-\f{V_{\sf ES}}{2\Phi_{\sf ES}},\label{s1}\\
\f{1}{a_{\sf ES}^2}\!\!\!&=&\!\!\!-(\gamma-1)\f{\rho_{\sf ES}}{\Phi_{\sf ES}}+\f{\sigma_{\sf ES}^2}{12\Phi_{\sf ES}^2}-\f{V_{\sf ES}}{2\Phi_{\sf ES}},\label{s2}\\\hspace{-1cm}
&&\hspace{-1cm}(4-3\gamma)\rho_{\sf ES}-\f{3\sigma_{\sf ES}^2}{8\Phi_{\sf ES}}-2V_{\sf ES}+\Phi_{\sf ES}V^{\prime}_{\sf ES}=0,\label{s3}
\eea
where $V_{{\sf ES}}=V(\Phi_{{\sf ES}})$ and $V'_{{\sf ES}}=V'(\Phi_{{\sf ES}})$. The system (\ref{s1})-(\ref{s3}) possesses fewer equations than the unknowns $a_{\sf ES},\rho_{\sf ES},\Phi_{\sf ES},\sigma_{\sf ES}$ (assuming the potential is given), thus the system is under-determined. However, we can eliminate $a_{\sf ES}$ from equations (\ref{s1}) and (\ref{s2}) and solve the resultant equation along with equation (\ref{s3}) for the remained {\sf ES} state parameters. In section (\ref{dynsysappes}) we will see that this solution corresponds to the fixed point, presented in (\ref{ES-sol}), of the dynamical system (\ref{dyn1})-(\ref{dyn4}) and we will discuss the physical properties of the solution in more detail via introducing some dimensionless variables and studying the corresponding dynamical system. Next we proceed to study the stability of {\sf ES} Universe against homogeneous and isotropic perturbations. We then consider small perturbations around the static solution given for the scale factor and the {\sf BD} scalar field. Let us define
\bea\label{perts}
a(t)=a_{{\sf ES}}[1+\zeta(t)],~~~~~\Phi(t)=\Phi_{{\sf ES}}[1+\xi(t)],
\eea
whereby we obtain
\be\label{pertsrs}
\rho=\rho_{{\sf ES}}+\delta\rho(\zeta)\simeq\rho_{{\sf ES}}(1-3\gamma\zeta),~~~~\sigma^2=\sigma_{{\sf ES}} ^2+\delta\sigma^2(\zeta)\simeq\sigma _{{\sf ES}}^2(1-6\zeta),
\ee
where $\zeta(t)\ll1$ and $\xi(t)\ll1$ are small perturbations. Substituting for perturbed values of the above variables into equations (\ref{ecbdfes2}) and (\ref{evoleqsf0}) we get
\bea
2\ddot{\zeta}+\ddot{\xi}+a_1\xi+{a_2}\zeta\!\!\!&=&\!\!\!0,\label{pert22}\\
\ddot{\xi}+{a_3}\zeta+{a_4}\xi\!\!\!&=&\!\!\!0,\label{pert23}
\eea
where
\bea\label{consabc}
{a_1}&=&\frac{(1-\gamma)\rho_{{\sf ES}}}{\Phi_{{\sf ES}}}+\frac{\sigma_{{\sf ES}}^2}{6\Phi_{{\sf ES}}^2}-\frac{V_{{\sf ES}}}{2\Phi_{{\sf ES}}},\\
{a_2}&=&\f{\sigma_{{\sf ES}}^2}{2\Phi_{{\sf ES}}^2}-\f{2}{a_{{\sf ES}}^2}-\f{3(\gamma-1)\gamma\rho_{{\sf ES}} }{\Phi_{{\sf ES}}},\\
{a_3}&=&\f{3(4-3\gamma)\gamma\rho_{{\sf ES}}}{2(\omega+3)\Phi_{{\sf ES}}}-\f{9\sigma_{{\sf ES}} ^2}{8(\omega+3)\Phi_{{\sf ES}}^2},\\
{a_4}&=&-\f{V_{{\sf ES}}^{\prime\prime}\Phi_{{\sf ES}}}{2(\omega+3)}-\f{3\sigma_{{\sf ES}}^2}{16(\omega+3)\Phi_{{\sf ES}}^2},
\eea
and we have neglected the terms which are of second order in $\zeta$ and $\xi$ and their derivatives. We also note that two constants that appear within (\ref{pert22}) and (\ref{pert23}) will be automatically vanished as a result of the background equations (\ref{s1})-(\ref{s3}). The system (\ref{pert22}) and (\ref{pert23}) is a coupled linear system of second-order ordinary differential equations which can be recast into equivalent form as
\begin{equation}
\begin{bmatrix} 
\ddot{\zeta}  \\
\ddot{\xi}   \\
\end{bmatrix} 
={\rm M} 
\begin{bmatrix} 
\zeta   \\
\xi  \\
\end{bmatrix},~~~~~~~
{\rm M}=\begin{bmatrix} 
\f{1}{2}(a_3-a_2) & \f{1}{2}(a_4-a_1)  \\
-a_{3} & -a_{4}  \\
\end{bmatrix}.
\end{equation}
The solutions of the above system can be obtained by seeking for the eigenvalues of the matrix ${\rm M}$. Denoting $\Omega_1^2$ and $\Omega_2^2$ as the eigenvalues of ${\rm M}$ we find the solutions and frequency of oscillations as
\begin{equation}\label{sols}
\begin{bmatrix} 
\zeta \\
\xi  \\
\end{bmatrix} 
=\sum_{j=1}^{4}
\begin{bmatrix} 
\f{{a_1}+\Omega_j^2}{{a_2}+2\Omega_j^2}{\sf C}_{j}  \\
-{\sf C}_{j}   \\
\end{bmatrix} 
e^{\Omega_{j}t},~~~~~~~~~~~\Omega_{j}=\pm\sqrt{\f{1}{4}\left[\mathcal{X}\pm \sqrt{\mathcal{Y}}\right]},
\end{equation}
where 
\bea
\mathcal{X}&=&{a_3}-2a_4-{a_2},\label{x}\\
\mathcal{Y}&=&{a_3}({a_3}+8{a_1}-4a_4)+4a_4(a_4-{a_2})+{a_2}({a_2}-2{a_3}),\label{y}
\eea
and ${\sf C}_j$ are arbitrary constants. If the real part of frequency of oscillations gets nonzero values, the perturbation modes will grow up during the dynamical evolution of the Universe. We then require, for the stability of the solution, that $\Omega_j$ be of pure imaginary type so that the amplitude of perturbations be non-growing. Therefore, the static solution is stable provided that 
\be\label{condsxy}
\mathcal{X}\pm \sqrt{\mathcal{Y}}<0.
\ee
In the next section we rewrite conditions (\ref{condsxy}) in terms of some dimensionless parameters and thus explore their possible restrictions on the free parameters of the model.
\section{Stability of {\sf ECBD} Theory Through The Dynamical System Approach}\label{dynsysappes}
In this section we present the dynamical system structure of the {\sf ECBD} theory and discuss the conditions under which the theory accepts a stable {\sf ES} solution. To this aim, we rewrite equations (\ref{ecbdfes1})-(\ref{conseeqs2}) in terms of the conformal time
\begin{align}
\eta=\int{\f{dt}{a(t)}},
\end{align}
 as follows 
\begin{align}
&\left(\f{a'}{a}+\f{\Phi'}{2\Phi}\right)^{2}+1=\f{\rho a^{2}}{3\Phi}-\f{\sigma^{2}a^{2}}{36\Phi^{2}}
+\f{\omega+3}{6}\left(\f{\Phi'}{\Phi}\right)^{2}-\f{V(\Phi)a^{2}}{6\Phi},\label{conf1}\\
&\f{2a''}{a}-\left(\f{a'}{a}\right)^{2}+\f{\Phi''}{\Phi}+\f{a'}{a}\f{\Phi'}{\Phi}+1=\f{(1-\gamma)\rho a^{2}}{\Phi}+\nonumber\\
&\f{\sigma^{2}a^{2}}{12\Phi^{2}}-\f{2\omega+3}{4}\left(\f{\Phi'}{\Phi}\right)^{2}-\f{V(\Phi)a^{2}}{2\Phi},\label{conf2}\\
&\f{\Phi''}{\Phi}+2\f{a'}{a}\f{\Phi'}{\Phi}=\nonumber\\
&\f{a^{2}}{2(\omega+3)}\left\{\f{1}{\Phi}\left[(4-3\gamma)\rho-\f{3\sigma^{2}}{8\Phi}\right]+\left(\f{dV(\Phi)}{d\Phi} -2\f{V(\Phi)}{\Phi}\right)\right\},\label{conf3}\\
&\rho'+3\f{a'}{a}\gamma\rho=0,\label{conf4}\\
&{\sigma^{2}}^{'}+6\f{a'}{a}\sigma^{2}=0\label{conf5}.
\end{align}
Now, equations (\ref{conf1})-(\ref{conf5}) can be recast to an equivalent form which are more useful to reconstruct the field equations of the theory as a dynamical system. To this aim, let us introduce the following dimensionless variables
\begin{align}\label{var}
&{\sf X}=\f{\Phi'}{6\Phi},~~~{\sf Y}=\f{a'}{a}+\f{\Phi'}{2\Phi},~~~{\sf Z}=\f{\sigma^{2}a^{2}}{36\Phi^{2}},~~~{\sf F}=\f{\rho a^{2}}{3\Phi},~~~{\sf N}=a^{2}\frac{dV(\Phi)}{d\Phi},\nonumber\\
&{\sf Q}=\f{V(\Phi)a^{2}}{6\Phi},~~~ r=\f{\Phi dV(\Phi)/d\Phi}{V(\Phi)},~~~ m=\f{\Phi d^{2}V(\Phi)/d\Phi^{2}}{dV(\Phi)/d\Phi}.
\end{align}
These variables parametrize the form of scalar field potential which is a useful representation in the dynamical system approach. This method has been already used in the literature in order to find a suitable description of cosmological solutions. For example, in ~\cite{Amendola} the authors have defined similar variables to parametrize $f({\sf R})$ function and in \cite{myselfart} this approach has been used to parametrize $f({\sf R,T})$ function. One can represent the behavior of potential $V(\Phi)$ in the plane of parameters $r$ and $m$ via eliminating $\Phi$ from the right hand side of their definitions. In Table~\ref{tab} some examples are provided. 
\begin{sidewaystable}[h!]
\begin{center}
\caption{The corresponding $r$-$m$ relation corresponding to some given potentials\textsuperscript{*}}
\begin{tabular}{l @{\hskip 0.4in} l@{\hskip 0.4in} l @{\hskip 0.4in}l}\hline\hline

Potential $V(\Phi)$&             $r$&               $m$&          $r$-$m$ relation\\[0.5 ex]
\hline
$\Phi^{\alpha}$&     $\alpha$&       $\alpha-1$&      $m=r-1$\\[0.5 ex]
$\lambda  \Phi ^{\alpha }+\eta  \Phi ^{\beta }$&     $\frac{\lambda  (\alpha -\beta ) \Phi ^{\alpha }}{\lambda  \Phi ^{\alpha }+\eta  \Phi ^{\beta }}+\beta$&       $\frac{\alpha  \lambda  (\alpha -\beta ) \Phi ^{\alpha }}{\alpha  \lambda  \Phi ^{\alpha }+\beta  \eta  \Phi ^{\beta }}+\beta-1$&      $m=-\frac{\alpha  \beta }{r}+\alpha +\beta -1$\\[0.5 ex]
$\Phi ^{\alpha } \log ^{\beta }(\lambda  \Phi )$&     $\frac{\beta }{\log (\lambda  \Phi )}+\alpha$&       $\frac{\alpha }{\alpha  \log (\lambda  \Phi )+\beta } +\frac{\beta -1}{\log (\lambda  \Phi )}+\alpha-1$&      $m=-\frac{(r-\alpha )^2}{\beta  r}+r-1$\\[0.5 ex]
$\Phi ^{\alpha } \exp \left(\lambda\Phi ^{\beta}\right)$&     $\alpha+\beta  \lambda  \Phi ^{\beta }$&       $\beta\left(\frac{\alpha\beta }{\beta  \lambda  \Phi ^{\beta }+\alpha}+\lambda \Phi ^{\beta }\right)+\beta+\alpha -1$&      $m=\frac{\alpha  \beta }{r}+r-(\beta+1)$\\[0.5 ex]
\hline\hline
\multicolumn{4}{l}{\footnotesize * Note that in each case the parameters $\alpha$, $\beta$, $\lambda$ and $\eta$ are constant whose dimensions can be determined by the fact that $r$ and $m$ are dimensionless.}
\end{tabular}
\label{tab}
\end{center}
\end{sidewaystable}

Therefore, the autonomous system of equations (\ref{conf1})-(\ref{conf5}) is attained as
\begin{align}
&{\sf X}'=-2{\sf X}{\sf Y}+\nonumber\\
&\frac{1}{4 (\omega +3)}\left\{(4-3 \gamma)\bigg[1-6(\omega +3) {\sf X}^{2}+{\sf Y}^{2}\bigg]+(2 r-3\gamma){\sf Q}-\frac{(6 \gamma+1) }{2}{\sf Z}\right\}\label{dyn1}\\
&{\sf Y}'=\left(1-\frac{3 \gamma }{2}\right)\left (1+{\sf Y}^{2}\right)+9 (\omega +3) (\gamma -2){\sf X}^2-\frac{3 \gamma}{2}{\sf Q}+3\left(1-\frac{\gamma }{2}\right) {\sf Z},\label{dyn2}\\
&{\sf Z}'=-4{\sf Z}{\sf Y}\label{dyn3},\\
&{\sf Q}'=2{\sf Q}\left[{\sf Y}+3(r-2){\sf X}\right],\label{dyn4}
\end{align}
where we have used equation (\ref{conf1}) in the following form
\begin{align}\label{conf6}
{\sf Y}^{2}+1={\sf F}-{\sf Z}+6(\omega+3){\sf X}^{2}-{\sf Q}.
\end{align}
Note that, in obtaining the system of equations (\ref{dyn1})-(\ref{dyn4}) the relation ${\sf N}=6r{\sf Q}$ has been used. The dynamical system described by differential equations (\ref{dyn1})-(\ref{dyn4}) admits nine fixed points which are presented in table \ref{tab2}. It is then observed that the only physically acceptable fixed point is ${\sf P}_1$, as the other points include imaginary values in the 4-dimensional phase-space constructed out of $({\sf X,Y,Z,Q})$ coordinates or correspond to non-vanishing time derivatives of the scale factor and the {\sf BD} scalar field. Hence, the only valid critical point of the system (\ref{dyn1})-(\ref{dyn4}) reads
\begin{align}\label{ES-sol}
{\sf X}_{{\sf ES}}=0,~~~~~{\sf Y}_{{\sf ES}}=0,~~~~~{\sf Z}_{\sf{ES}}=\frac{4\left[3 \gamma +(2-3 \gamma ) r_{{\sf ES}}\right]}{3 \left[13 \gamma +4 (\gamma -2) r_{{\sf ES}}\right]},~~~~~
{\sf Q}_{{\sf ES}}=\frac{(50-51 \gamma )}{3 \left[13 \gamma +4 (\gamma -2) r_{{\sf ES}}\right]}.
\end{align}
\begin{center}
	\begin{table}[ht]
		\centering
		\caption{The fixed point solutions of the autonomous system (\ref{dyn1})-(\ref{dyn4}).}
		\begin{tabular}{l @{\hskip 0.7in} l}\hline\hline\\[-1.5 ex]
			Point     &Coordinates $({\sf X,Y,Q,Z})$\\[0.4 ex]
			\hline\\[-1 ex]
			${\sf P}_{1}$&$\left(0,0,\frac{(50-51 \gamma )}{3 (13 \gamma +4 (\gamma -2) r)},\frac{4 (3 \gamma +(2-3 \gamma ) r)}{3 (13 \gamma +4 (\gamma -2) r)}\right)$\\[2 ex]
			${\sf P}_{2}$&$\left(0,-i,0,0\right)$\\[2 ex]
			${\sf P}_{3}$&$\left(0,i,0,0\right)$\\[2 ex]
			${\sf P}_{4}$&$\left(-\frac{1}{3\sqrt{2 (1 + \omega) + r(4 - r) }},\frac{r-2}{3\sqrt{2 (1 + \omega) + r(4 - r) }},\frac{4(3+\omega)}{3 [(-4 + r) r - 2 (1 + \omega)]},0\right)$\\[2 ex]
			${\sf P}_{5}$&$\left(\frac{1}{3\sqrt{2 (1 + \omega) + r(4 - r)}},\frac{2-r}{3\sqrt{2 (1 + \omega) + r(4 - r)}},\frac{4(3+\omega)}{3 [(-4 + r) r - 2 (1 + \omega)]},0\right)$\\[2 ex]
			${\sf P}_{6}$&$\left(-\frac{\sqrt{51 \gamma-50}}{9\sqrt{2(2-\gamma) (3 + \omega)}}i,0,0,\frac{8 (3 \gamma -4) }{27 (\gamma -2)}\right)$\\[2 ex]
			${\sf P}_{7}$&$\left(\frac{\sqrt{51 \gamma-50}}{9\sqrt{2(2-\gamma) (3 + \omega)}}i,0,0,\frac{8 (3 \gamma -4) }{27 (\gamma -2)}\right)$\\[2 ex]
			${\sf P}_{8}$&$\left(\frac{(2-3 \gamma )}{3 \sqrt{\gamma  (3 \gamma -8) (2 \omega +3)+8 (\omega +1)}},\frac{(4-3 \gamma )}{\sqrt{\gamma  (3 \gamma -8) (2 \omega +3)+8 (\omega +1)}},0,0\right)$\\[2 ex]
			${\sf P}_{9}$&$\left(\frac{(3 \gamma-2 )}{3 \sqrt{\gamma  (3 \gamma -8) (2 \omega +3)+8 (\omega +1)}},\frac{(3 \gamma -4)}{\sqrt{\gamma  (3 \gamma -8) (2 \omega +3)+8 (\omega +1)}},0,0\right)$\\[2 ex]
			\hline\hline
		\end{tabular}
		\label{tab2}
	\end{table}
\end{center}
The critical values of variables ${\sf X}$ and ${\sf Y}$, are located at the center of $({\sf X,Y})$ plane while those of spin contribution ${\sf Z}$ and potential term ${\sf Q}$ depend on the {\sf EoS} parameter of the perfect fluid as well as the critical value of $r$, only. Note that the parameter $r$ must be also a constant at the fixed point solution. To show this, it is easy to obtain the equation for the evolution of $r$ given as
\begin{align}\label{rdot}
r'=6r{\sf X}(1-r+m).
\end{align}
Therefore, for ${\sf X}_{{\sf ES}}=0$ we have $r'=0$ and thus $r=r_{{\sf ES}}=r(\Phi_{{\sf ES}})$. Using the dynamical variables (\ref{var}), for a specified form of the scalar field potential, namely, $V(\Phi)={\sf A}\Phi^{n}$ together with taking the square of spin density as a free parameter and using the fixed point solution (\ref{ES-sol}) we find the critical values $\Phi_{{\sf ES}}$, $a_{{\sf ES}}$ and $\rho_{{\sf ES}}$ as
\begin{align}
&a_{{\sf ES}}=6\left(\frac{6{\sf A}{\sf Z}_{{\sf ES}}}{{\sf \sigma}^{2}_{{\sf ES}}\sf{Q}_{\sf ES}}\right)^{-\frac{1}{n+1}}\sqrt{\frac{{\sf Z}_{{\sf ES}}}{{\sf \sigma}^{2}_{{\sf ES}}}}.\label{p-a}\\
&\Phi_{{\sf ES}}=\left(\frac{6{\sf A}{\sf Z}_{{\sf ES}}}{{\sf \sigma}^{2}_{{\sf ES}}\sf{Q}_{\sf ES}}\right)^{-\frac{1}{n+1}},\label{p-phi}\\
&\rho_{{\sf ES}}=\frac{F_{\sf ES} {\sf \sigma}^{2}_{{\sf ES}}}{12 {\sf Z}_{{\sf ES}}}\left(\frac{6{\sf A}{\sf Z}_{{\sf ES}}}{{\sf \sigma}^{2}_{{\sf ES}}\sf{Q}_{\sf ES}}\right)^{\frac{1}{n+1}},\label{p-rho}
\end{align}
where ${\sf A}$ is a constant. From (\ref{p-a}) we observe that non-vanishing size of initial scale factor of {\sf ES} Universe depends on the {\sf EoS} parameter along with the spin density and the the value of $r$ parameter at the equilibrium point. Note that, in (\ref{p-a})-(\ref{p-rho}) parameters ${\sf Z}$, ${\sf F}$ and ${\sf Q}$ are functions of $\gamma$ and $r$ and also using the definition $r$ in $(\ref{var})$ we get $r=n$. For more complicated forms of the scalar field potential we have $r=r(\Phi_{{\sf ES}})$, (see Table \ref{tab}) and thus the problem of obtaining the critical values $\Phi_{{\sf ES}}$, $a_{{\sf ES}}$ and $\rho_{{\sf ES}}$ may not be an easy task. For a general form of potential one may find some restrictions on free parameters of the theory by applying the conditions $\rho_{{\sf ES}}>0$, $a_{{\sf ES}}>0$ and $\sigma_{{\sf ES}}>0$. \par
\begin{figure}
	\begin{center}
		\includegraphics[scale=0.22]{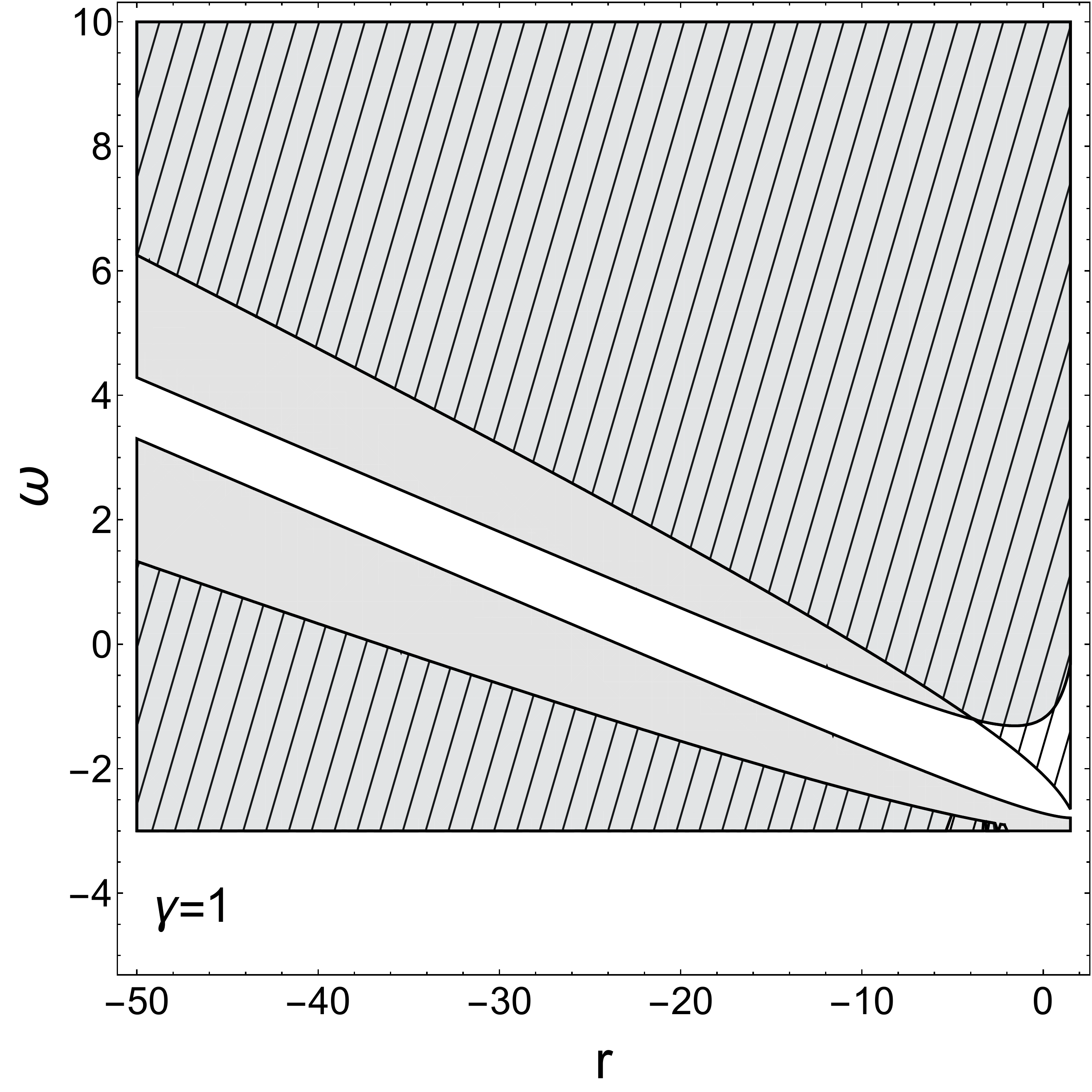}
		\includegraphics[scale=0.22]{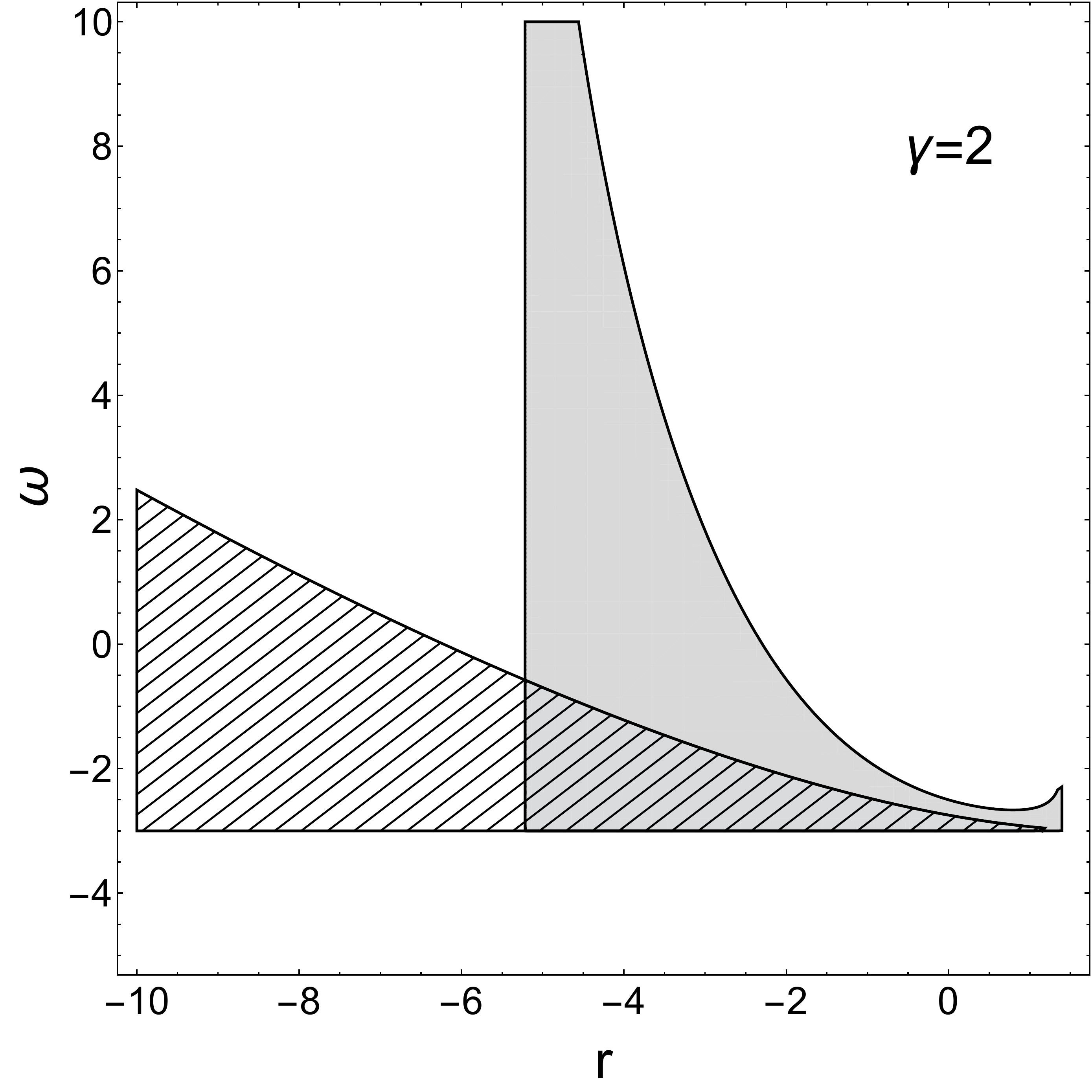}
		\includegraphics[scale=0.22]{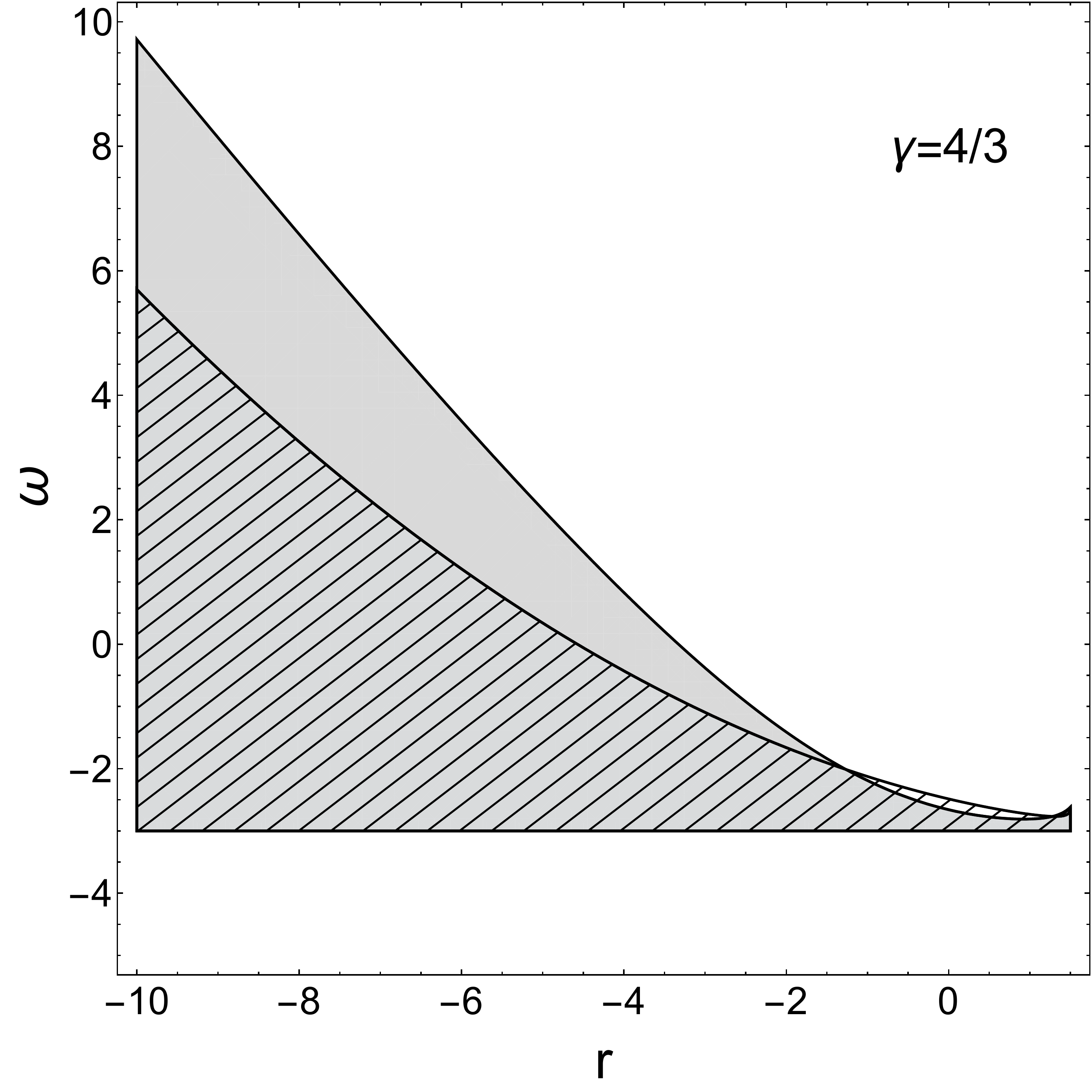}
		\caption{The space parameter for three different {\sf EoS} parameters i.e., dust (left panel), stiff matter (right panel) and radiation (lower panel). The shaded regions show the allowed values of $(\omega,r)$ parameters for which the fixed point is a center equilibrium. The gray regions show the allowed values for these parameters (for power-law potential) for which the solution is stable against small perturbations. The white regions are not allowed.}\label{fig1}
	\end{center}
\end{figure}

 From the definition of variables (\ref{var}) we require that ${\sf Z}_{{\sf ES}}>0$ from which we obtain the following conditions
\begin{align}
&r\leq \frac{3}{2},~~~~~~~~~ 1\leq\gamma\leq2,\\
&\frac{3}{2}<r<3,~~~~~~~~~ 1<\gamma <\frac{2 r}{3 r-3},\\
&r>\frac{13}{4},~~~~~~~~~  1<\gamma <\frac{8 r}{4 r+13}.
\end{align}
The eigenvalues of the system (\ref{dyn1})-(\ref{dyn4}) are some complicated functions of $\omega$, $r$ and $\gamma$, which can be written as follows
\begin{align}\label{eig}
\lambda_{1,..,4}=\pm\frac{\sqrt{3}}{2 (\omega +3)}\left[f(\gamma,\omega,r)\pm\sqrt{g(\gamma,\omega,r)}\right]^{\f{1}{2}},
\end{align}
where we have defined
\begin{align}
f(\gamma,\omega,r)=4 (\gamma -2) (\omega +3)^2 {\sf Z}_{{\sf ES}}- (\omega +3) \left[2 \gamma  \omega -2 r^2+(3 \gamma +4) r\right]{\sf Q}_{{\sf ES}},
\end{align}
and 
\begin{align}
&g(\gamma,\omega,r)=(\omega +3)^2\Bigg\{{\sf Q}_{{\sf ES}}^2 \Big[2 \gamma  \omega -2 r^2+(3 \gamma +4) r\Big]^2\nonumber\\
&+16 (\gamma-2)^2 (\omega +3)^2 {\sf Z}_{{\sf ES}}^2-8(\omega +3){\sf Q}_{{\sf ES}} {\sf Z}_{{\sf ES}}\times\nonumber\\
& \bigg[-26 \gamma+2\gamma\omega (\gamma-2)+r \Big(3 \gamma  (\gamma +1)+2r(\gamma -2) +8\Big)\bigg]\Bigg\}.
\end{align}
In order that the fixed point (\ref{ES-sol}) represents an {\sf ES} solution, the above four eigenvalues must get pure imaginary values, simultaneously. That is, the {\sf ES} state dynamically corresponds to a center equilibrium point from which small departures result in oscillations around that point instead of exponential deviation from it. In such a situation the Universe stays (oscillates) in the neighborhood of the {\sf ES} solution indefinitely. The shaded region in Fig.~(\ref{fig1}) presents the allowed values for the pair $(\omega,r)$ where the critical point (\ref{ES-sol}) is a center equilibrium point for three specific {\sf EoS} parameters.
\par
In order to study the behavior of small perturbations we begin by rewriting the constants ${a}_1-{a}_4$ in terms of variables (\ref{var}), as follows
\begin{align}
&{a}_1=3 [{\sf F}(1-\gamma)-{\sf Q}+{\sf Z}],\label{a2}\\
&{a}_2=-9 (\gamma -1) {\sf F}+18 {\sf Z}-2,\label{b2}\\
&{a}_3=-\frac{9 \left[(3 \gamma -4) {\sf F}+9 {\sf Z}\right]}{2(\omega +3)},\label{c2}\\
&{a}_4=\frac{3 (4 m {\sf Q} r-9 {\sf Z})}{4 (\omega +3)},\label{d2}
\end{align}
Substituting relations (\ref{a2})-(\ref{d2}) into definitions (\ref{x})-(\ref{y}) together with using (\ref{ES-sol}) gives the values of $\mathcal{X}$ and $\mathcal{Y}$ in terms of the four parameters ($\gamma$, $\omega$, $r$) and $m$. Therefore, in cases where the relation $m=m(r)$ is specified for a given potential, the space of free parameters reduces to a three dimensional one constructed out of $(\gamma,\omega,r)$. However, conditions (\ref{condsxy}), could generally set some different restrictions on model parameters in comparison with those obtained from (\ref{eig}). Nevertheless, it is still possible to obtain a space of parameters for which the {\sf ES} state is stable and conditions (\ref{condsxy}) are fulfilled. In such a situation the Universe oscillates about an initial {\sf ES} solution and despite of nearly slight departures from this state due to the homogeneous perturbations, the Universe continues its dynamical evolution around the stable state without ever collapsing to zero radius or expansion. In the case of power law potential we have $m=r-1$ (see Table~\ref{tab}) and thus the definitions (\ref{x})-(\ref{y}) yield
\begin{align}
&\mathcal{X}=\left\{(\omega +3) [13 \gamma +4 (\gamma -2) r]\right\}^{-1}\Big{\{}2 r [87 \gamma +(51 \gamma -50) r-58]\nonumber\\
&+8 \omega  [13 \gamma +(4 \gamma -2) r]-3 [7 \gamma +50 (\omega +1)]\Big{\}},\\
&\mathcal{Y}=\left\{(\omega +3)^2 [13 \gamma +4 (\gamma -2) r]^2\right\}^{-1}\Bigg{\{}4r^4 (50-51 \gamma )^2-229491 \gamma ^2\nonumber\\
&+4\omega ^2 (75-52 \gamma )^2+2580 \omega\gamma (15-26 \gamma ) +4500 (51 \gamma +10 \omega +5)\nonumber\\
&+4 r^2 \Big{[}-42723 \gamma ^2+64 (1-2 \gamma )^2 \omega ^2+2\omega[3 \gamma  (724 \gamma -437)-566] \nonumber\\
&+90210 \gamma -46688\Big{]}-32r^3 (51 \gamma -50)[\gamma  (4 \omega -27)-2 \omega +28]\nonumber\\
&+4 r \Big{[}\gamma ^2 [4 \omega  (416 \omega +6549)+87129]-2 \gamma[\omega  (1616 \omega +27353)+83616]
&+300[\omega  (4 \omega +85)+253]\Big{]}\Bigg{\}}.
\end{align}
The gray zones in Fig.~(\ref{fig1}) present the allowed regions in ($\omega,r$) plane for which conditions (\ref{condsxy}) are satisfied and thus for any point picked up from these regions, the static state is stable against small perturbations. It is notable that the study of behavior of perturbations is carried out for a given form of potential (power law), while, there is less constraint for specifying the form of potential in the dynamical system approach which can be utilized for the potentials with constant $r$ or those that satisfy the relation $m=r-1$, see equation (\ref{rdot}). Thus, the observed difference between the gray and shaded regions of Fig. (\ref{fig1}) can be related to this issue. Nevertheless, each approach accepts its own stability space parameter and for a point picked up from uncommon zones the two approaches may not necessarily lead to the same results. For example as shown in Fig.~(\ref{figrev}) we have plotted the evolution of scale factor and the scalar field for $\gamma=2$ and ($\omega,r$) located in the white shaded region of Fig. (\ref{fig1}). We therefore observe that the Universe undergoes oscillations around its stable state while this scenario cannot happen from the viewpoint of perturbation method as the chosen point is out of the gray region.
\par
\begin{figure}
	\begin{center}
		\includegraphics[scale=0.27]{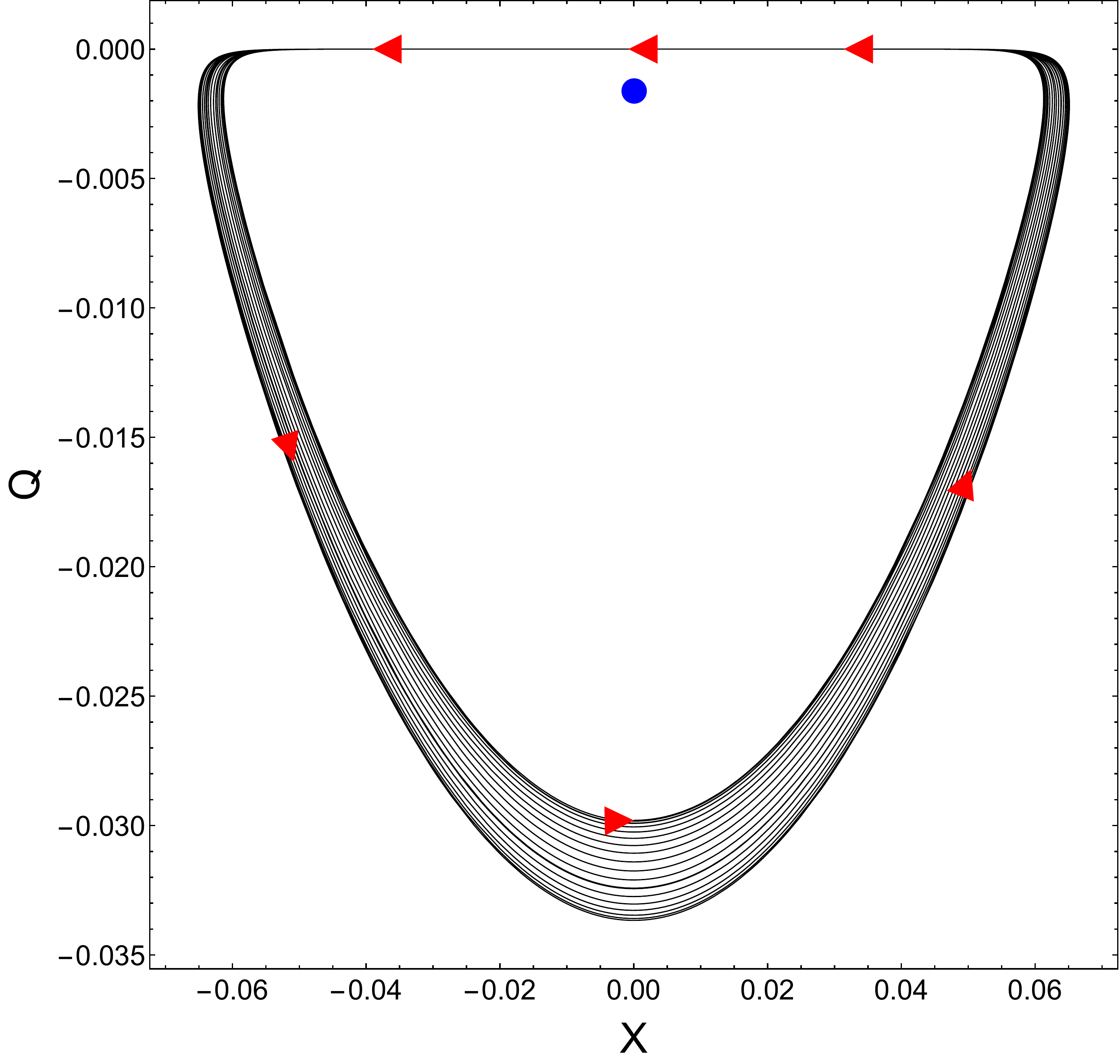}
		\includegraphics[scale=0.26]{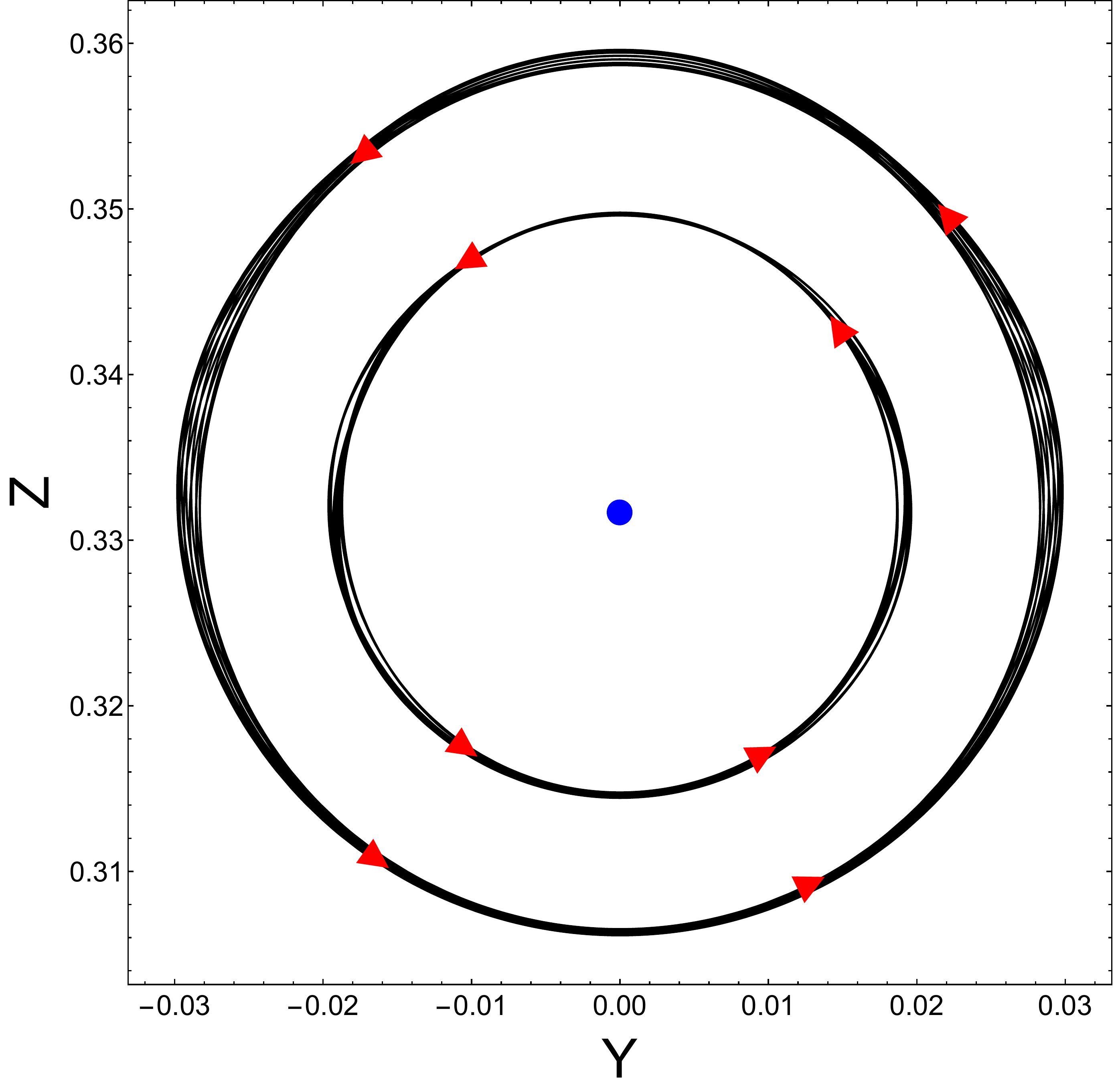}
		\includegraphics[scale=0.35]{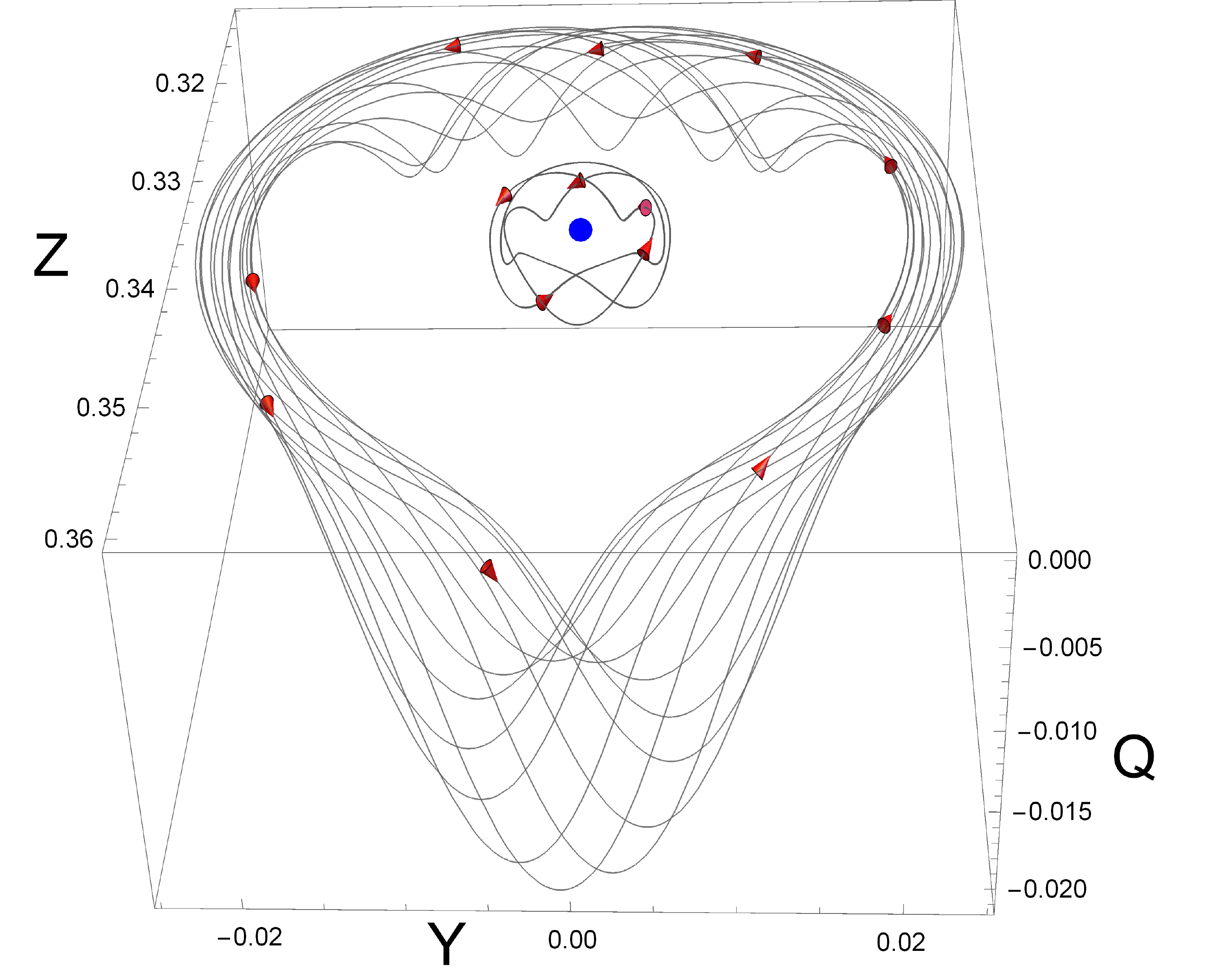}
		\caption{Dynamical evolution of the Universe dominated by dust $(\gamma=1)$. The initial values for numerical integration of the system (\ref{dyn1})-(\ref{dyn4}) along with model parameters have been set as, $\omega=-2$, $r=-48$, ${\sf X}_0={\sf Y}_0=0$, ${\sf Z}_0=0.3340$, ${\sf Q}_0=-0.0298$ for the left panel and ${\sf Z}_0=0.309$, ${\sf Q}_0=-0.0117$ for outer orbits and ${\sf Z}_0=0.3496$, ${\sf Q}_0=-0.003$ for inner orbits of the right panel. The blue point exhibits the location of fixed points given by ${\sf X}_{\sf ES}=0$, ${\sf Q}_{\sf ES}=-0.00162$ (left plot) and ${\sf Y}_{\sf ES}=0$, ${\sf Z}_{\sf ES}=0.3317$ (right plot). The lower panel represents numerical simulation of the solution in 3D subspace $({\sf Z,Q,Y})$ for the same model parameters as the upper plots. The initial values for numerical integration has been chosen as ${\sf X}_0={\sf Y}_0=0$, ${\sf Z}_0=0.36$, ${\sf Q}_0=-0.0213$ (outer orbits) and ${\sf Z}_0=0.337$, ${\sf Q}_0=-0.00394$ (inner orbits).}\label{fig2}
	\end{center}
\end{figure}

\begin{figure}
	\begin{center}
		\includegraphics[scale=0.27]{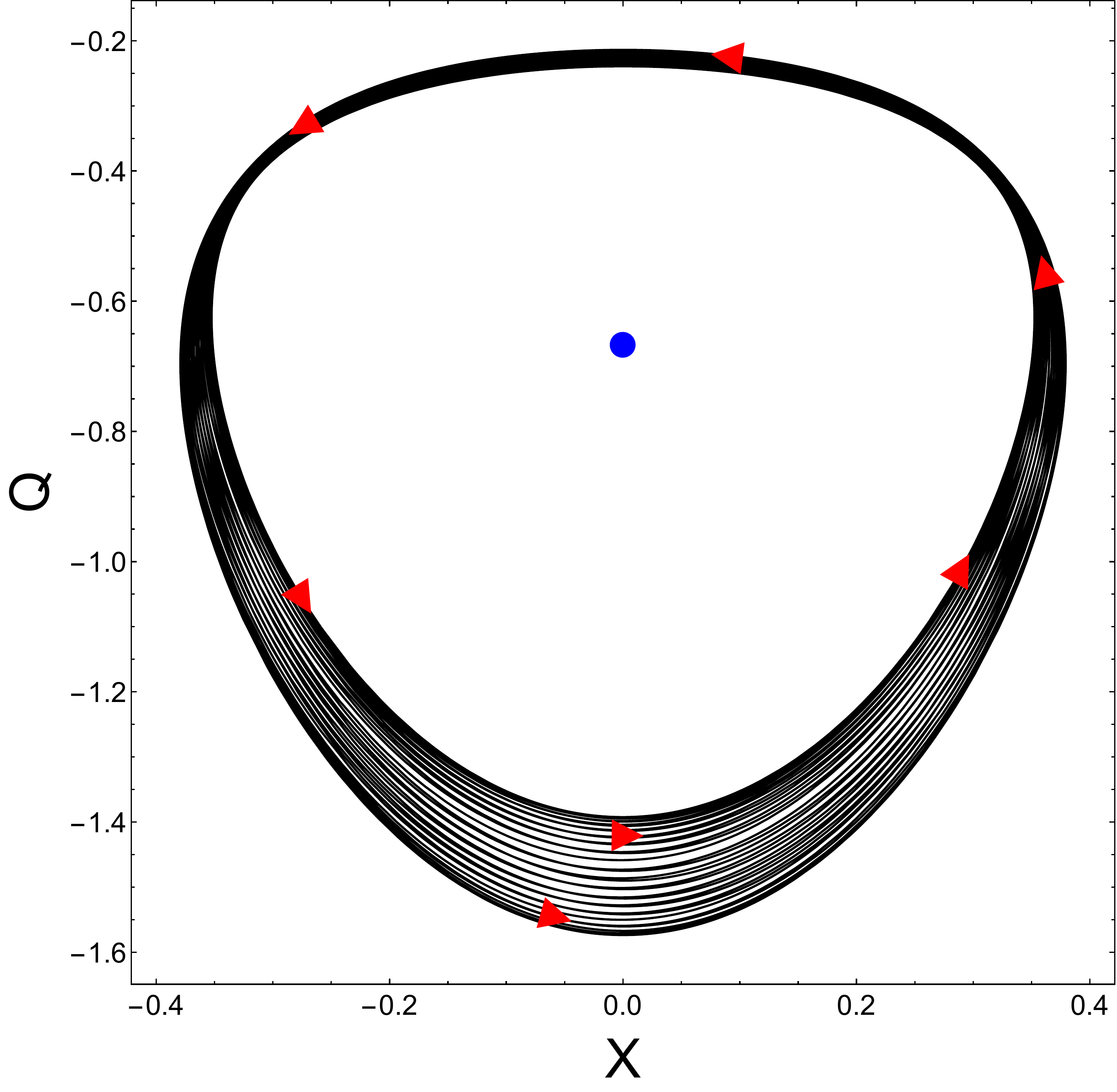}
		\includegraphics[scale=0.27]{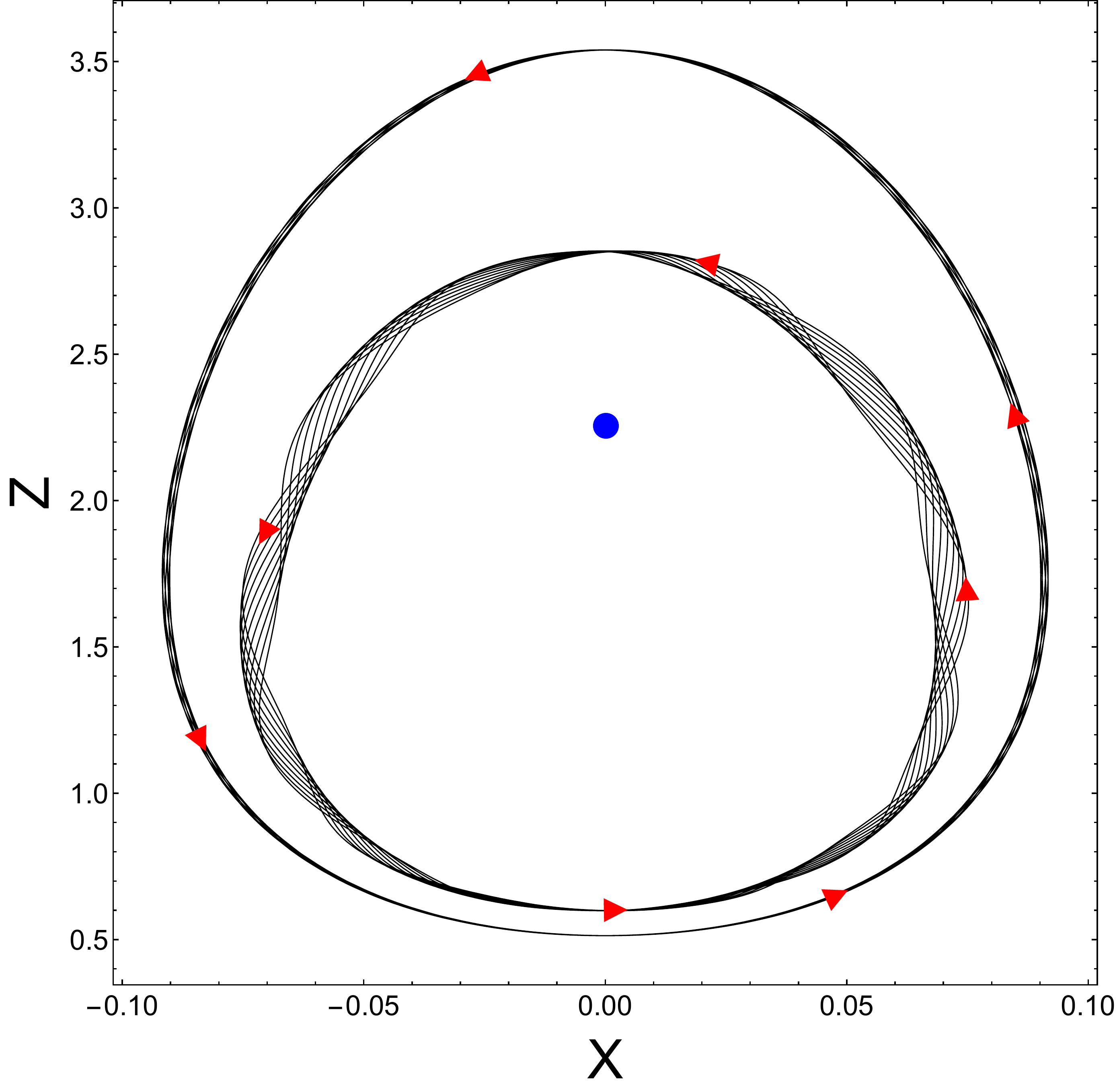}
		\includegraphics[scale=0.28]{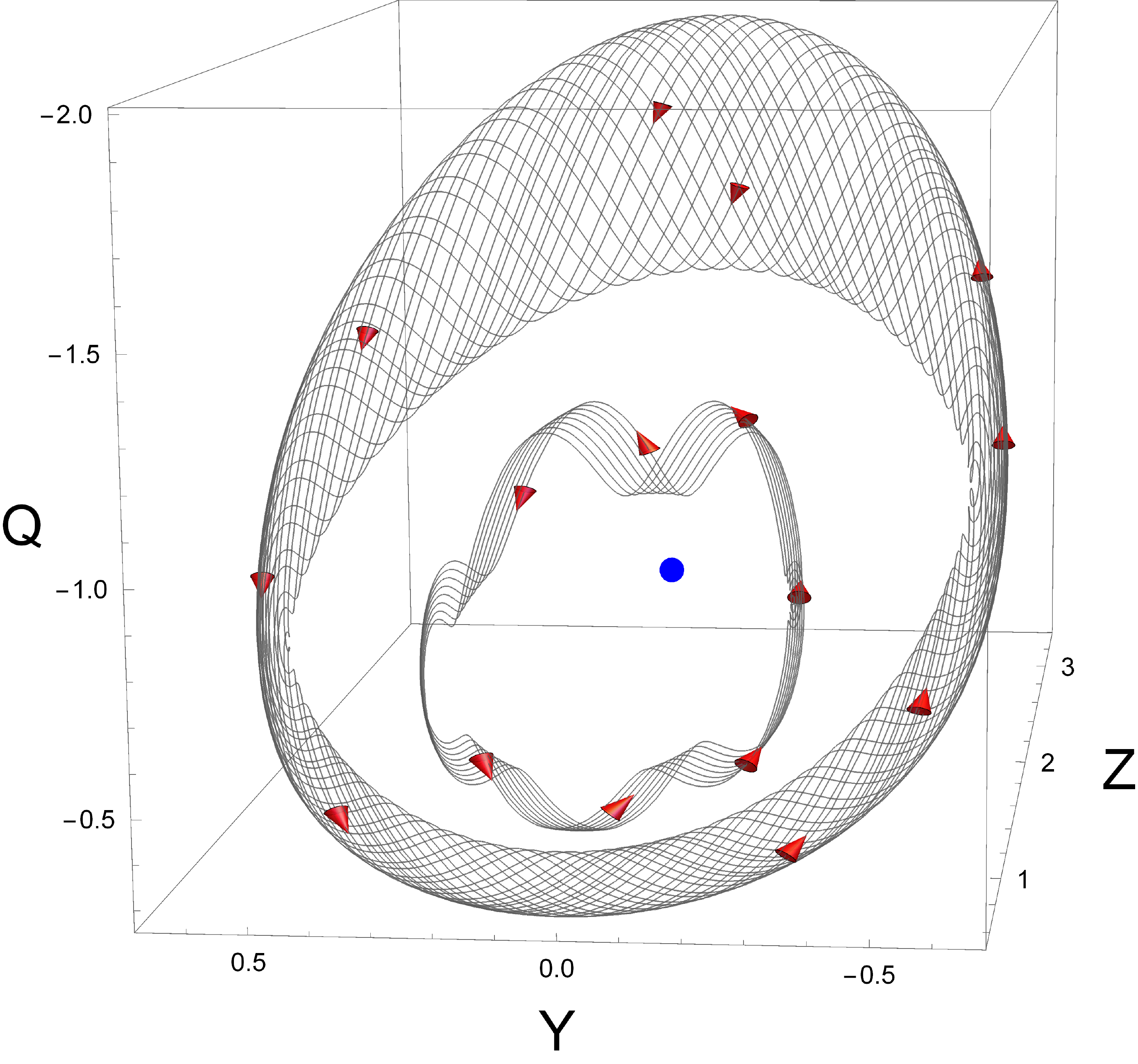}
		\caption{Dynamical evolution of the Universe dominated by a stiff fluid $(\gamma=2)$. The initial values for numerical integration of the system (\ref{dyn1})-(\ref{dyn4}) along with model parameters have been set as, $\omega=-2.6$, $r=-4.5$, ${\sf X}_0={\sf Y}_0=0$, ${\sf Z}_0=1.156$, ${\sf Q}_0=-0.214$ (left panel) and ${\sf Z}_0=3.539$, ${\sf Q}_0=-1.7087$ for outer orbits and ${\sf Z}_0=2.8511$, ${\sf Q}_0=-1.3890$ for inner orbits of the right panel. The location of fixed points are given by ${\sf X}_{\sf ES}=0$, ${\sf Q}_{\sf ES}=-0.666$ (left plot) and ${\sf Y}_{\sf ES}=0$, ${\sf Z}_{\sf ES}=2.2564$ (right plot). The lower panel represents numerical simulation of the solution in 3D subspace $({\sf Z,Y,Q})$ for the same model parameters as the upper plots. The initial values for numerical integration has been chosen as ${\sf X}_0={\sf Y}_0=0$, ${\sf Z}_0=3.339$, ${\sf Q}_0=-1.979$ (outer orbits) and ${\sf Z}_0=2.05$, ${\sf Q}_0=-1.17$ (inner orbits).}\label{fig3}
	\end{center}
\end{figure}

\begin{figure}
	\begin{center}
		\includegraphics[scale=0.27]{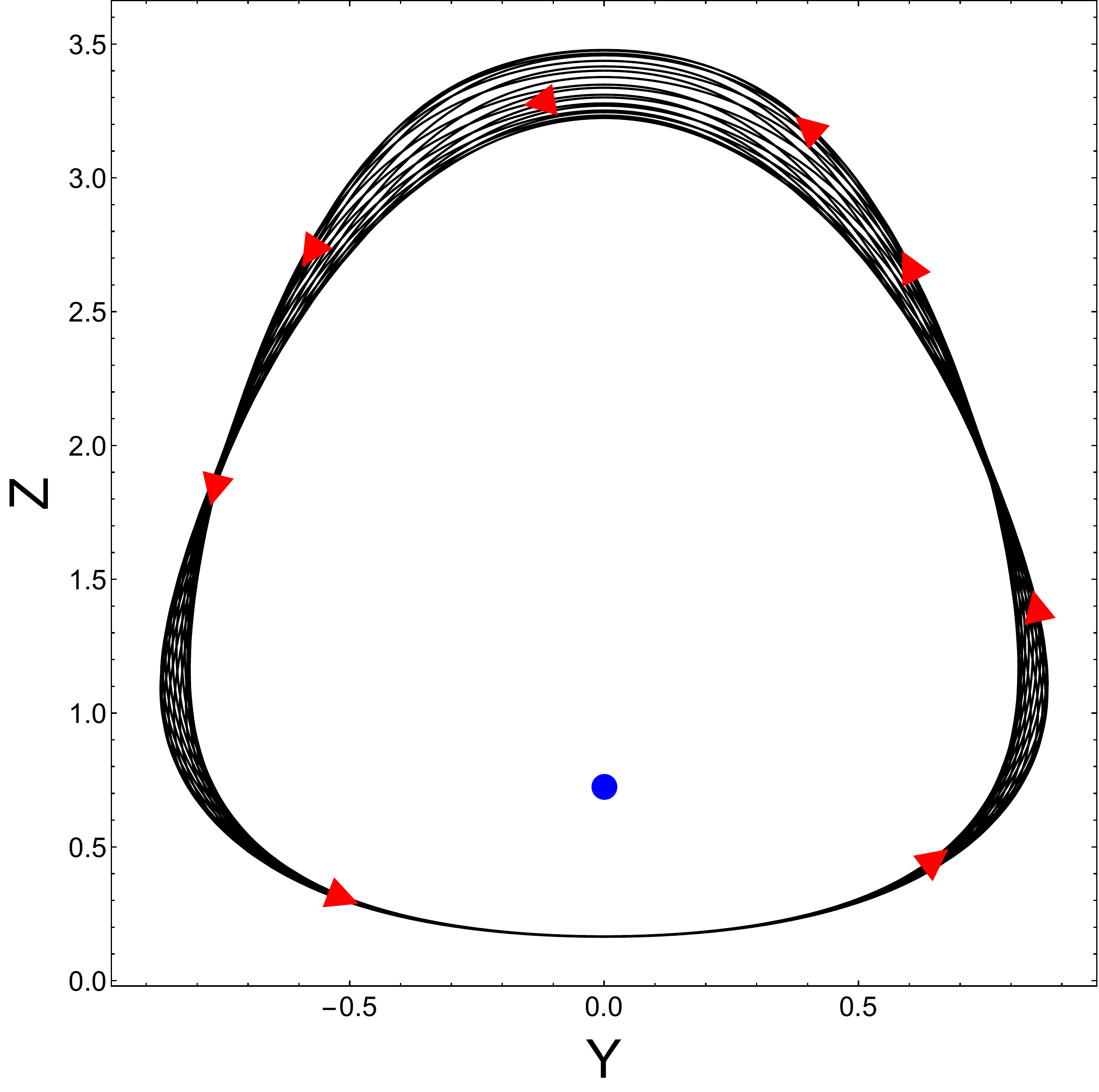}
		\includegraphics[scale=0.27]{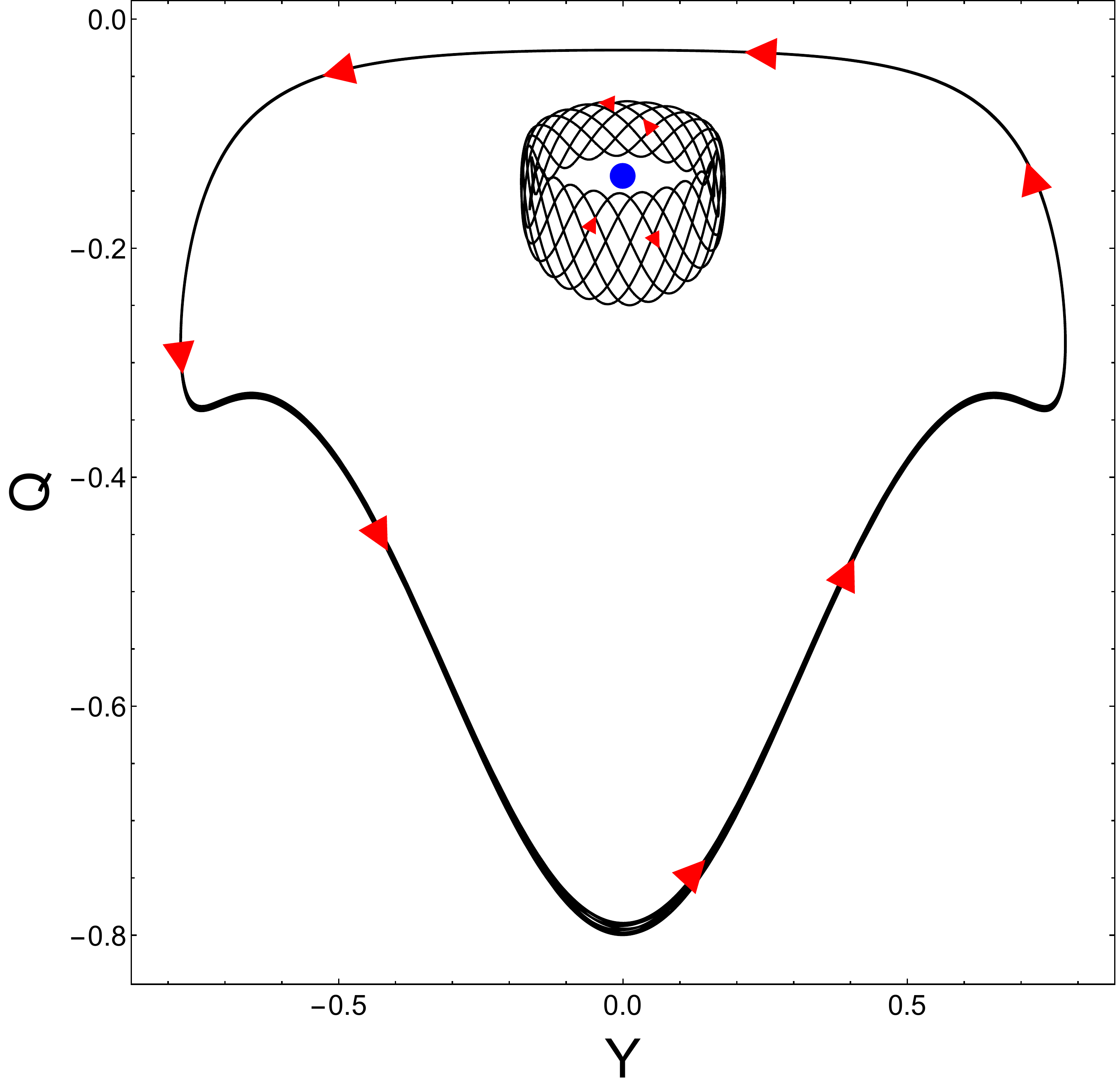}
		\includegraphics[scale=0.27]{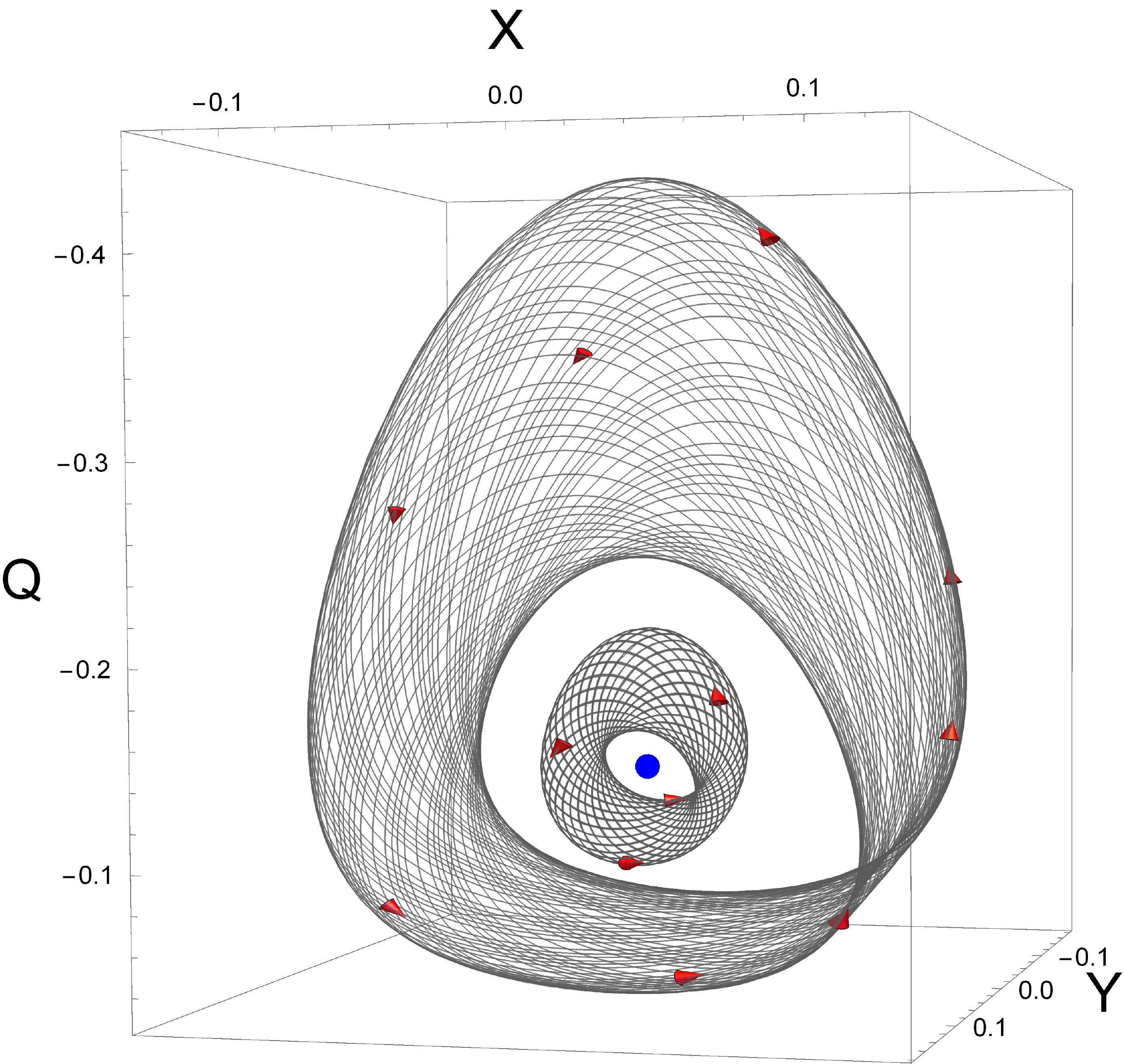}
		\caption{Dynamical evolution of the Universe dominated by radiation $(\gamma=\f{4}{3})$. The initial values for numerical integration of the system (\ref{dyn1})-(\ref{dyn4}) along with model parameters have been set as, $\omega=-2$, $r=-10$, ${\sf X}_0={\sf Y}_0=0$, ${\sf Z}_0=3.11$, ${\sf Q}_0=-0.5$ (left panel) and ${\sf Z}_0=3.01$, ${\sf Q}_0=-0.8$ for outer orbits and ${\sf Z}_0=1.012$, ${\sf Q}_0=-0.152$ for outer orbits of the right panel. The location of fixed points are given by ${\sf Z}_{\sf ES}=0.723$, ${\sf Y}_{\sf ES}=0$ (left plot) and ${\sf Y}_{\sf ES}=0$, ${\sf Q}_{\sf ES}=-0.136$ (right plot). The lower panel represents numerical simulation of the solution in 3D subspace $({\sf X,Y,Q})$ for the same model parameters as the upper plots. The initial values for numerical integration has been chosen as ${\sf X}_0={\sf Y}_0=0$, ${\sf Z}_0=1.01$, ${\sf Q}_0=-0.45$ (outer orbits) and ${\sf Z}_0=0.83$, ${\sf Q}_0=-0.1163$ (inner orbits).}\label{fig4}
	\end{center}
\end{figure}

\begin{figure}
\begin{center}
\includegraphics[scale=0.2]{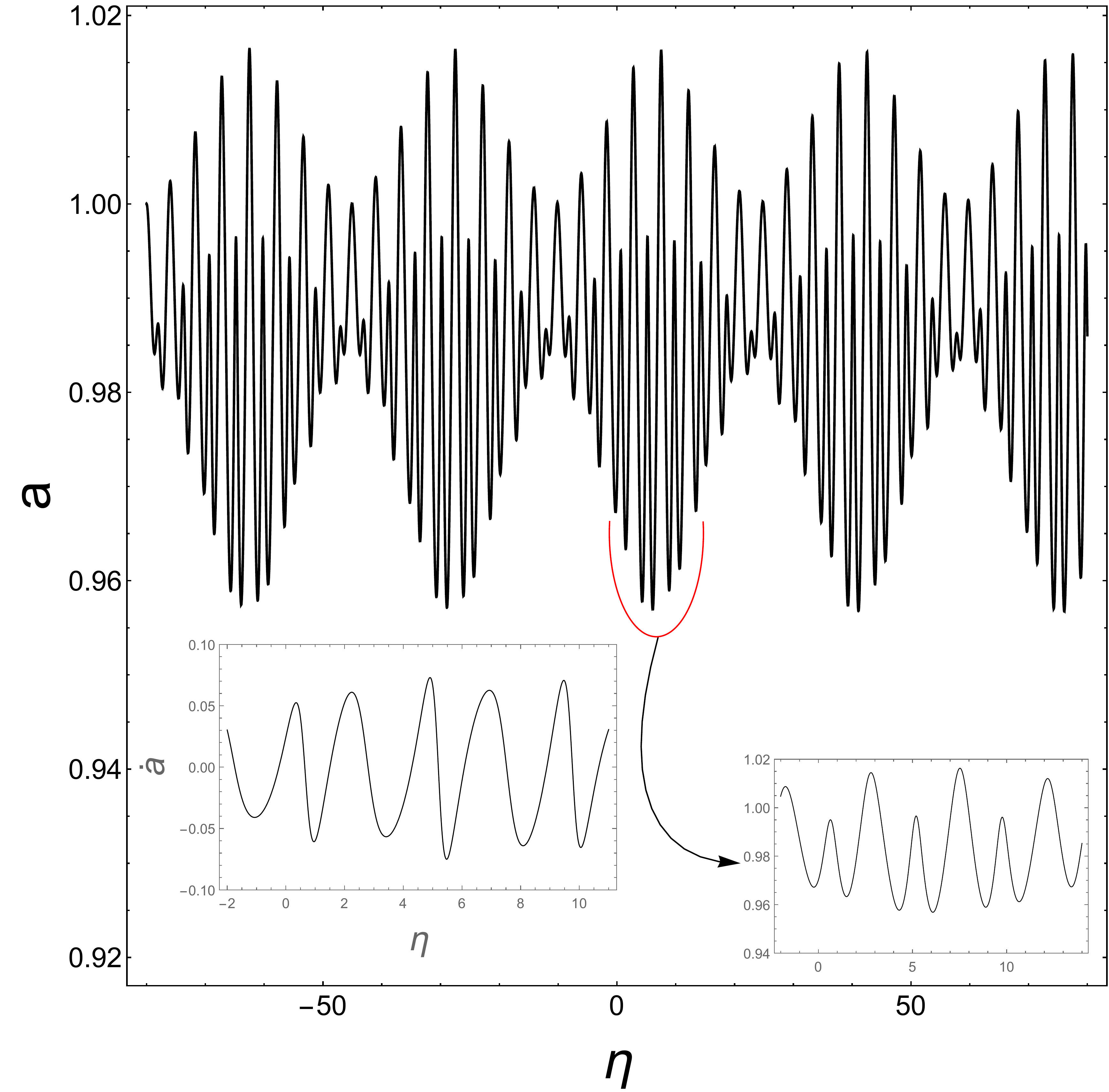}
\includegraphics[scale=0.2]{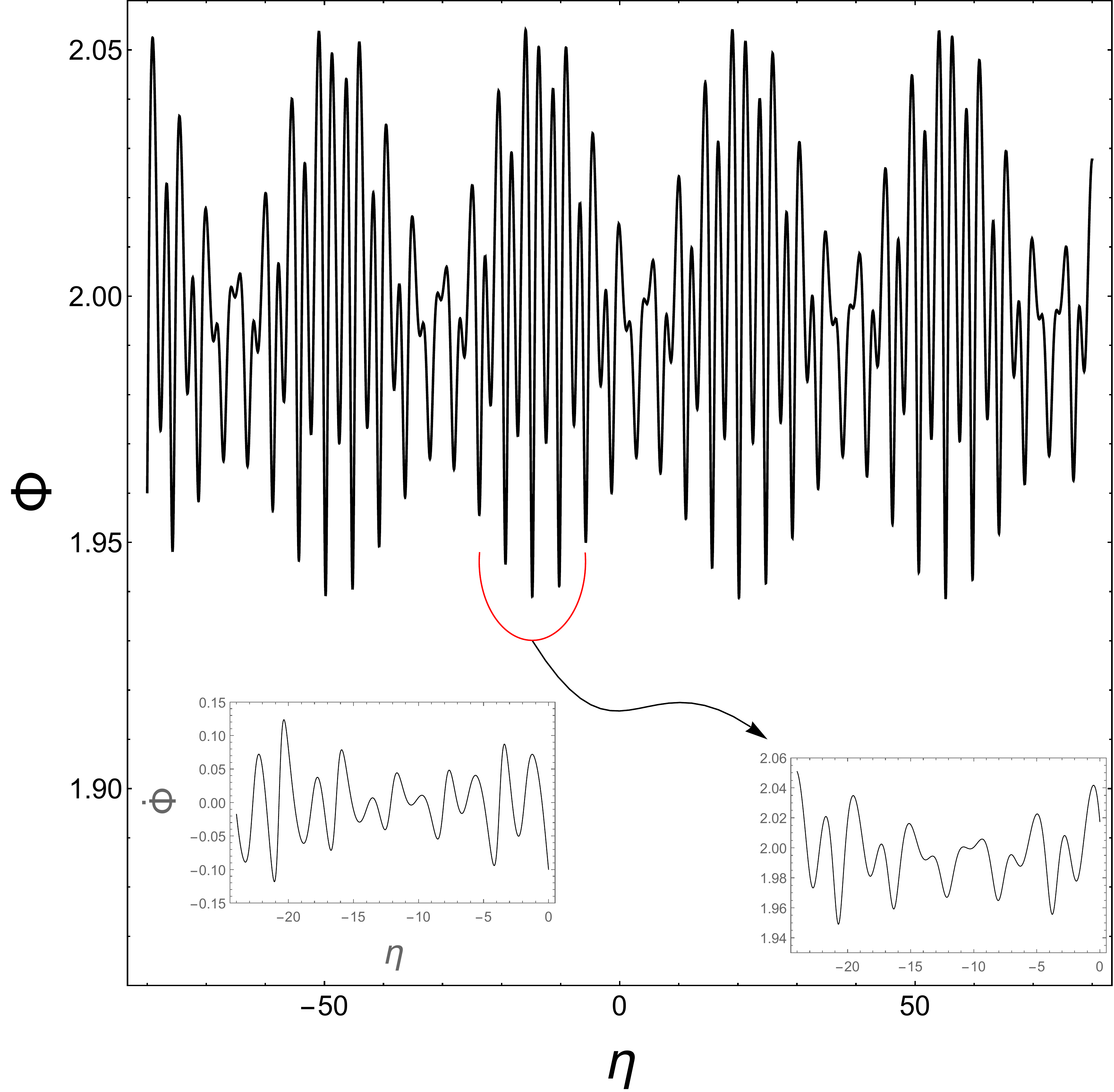}
\caption{Dynamical evolution of scale factor (left panel) and the {\sf BD} scalar field (right panel) for $\gamma=1$, $\omega=-2$ and $r=-34$. The initial values for dynamical variables has been set as ${\sf X}_0={\sf Y}_0=0$, ${\sf Z}_0=0.3117$ and ${\sf Q}_0=-0.00152$. The right insets show the behavior of $a$ and $\Phi$ within a shorter time interval and the left ones are plotted for time derivatives of these quantities.}\label{figaphi1}
\end{center}
\end{figure}

\begin{figure}
	\begin{center}
		\includegraphics[scale=0.2]{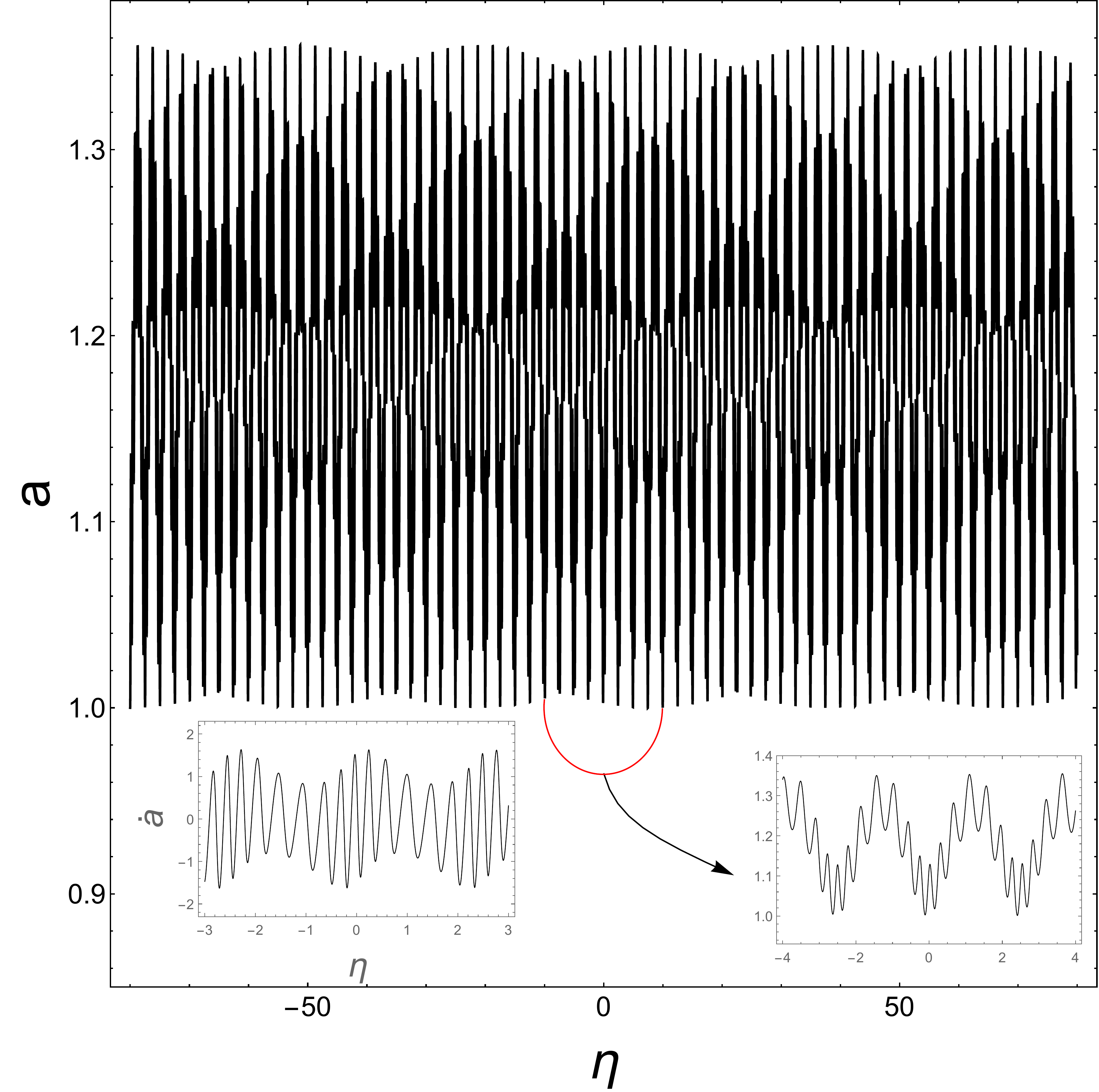}
		\includegraphics[scale=0.2]{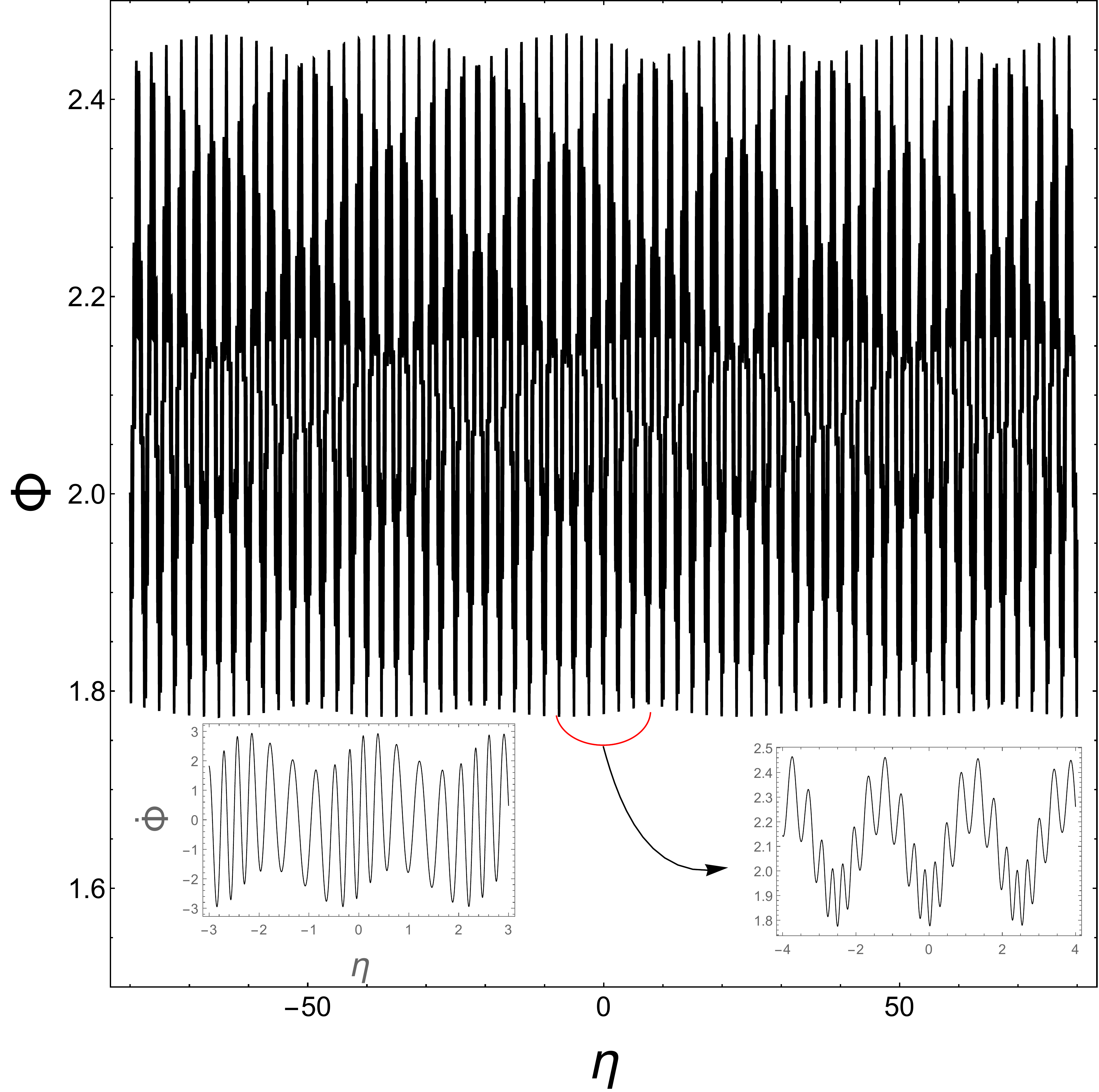}
		\caption{Dynamical evolution of scale factor (left panel) and the {\sf BD} scalar field (right panel) for $\gamma=2$, $\omega=-2.8$ and $r=-3$. The initial values for dynamical variables has been set as ${\sf X}_0={\sf Y}_0=0$, ${\sf Z}_0=2.116$ and ${\sf Q}_0=-0.66$. The right insets show the behavior of $a$ and $\Phi$ within a shorter time interval and the left ones are plotted for time derivatives of these quantities.}\label{figaphi2}
	\end{center}
\end{figure}

\begin{figure}
	\begin{center}
		\includegraphics[scale=0.2]{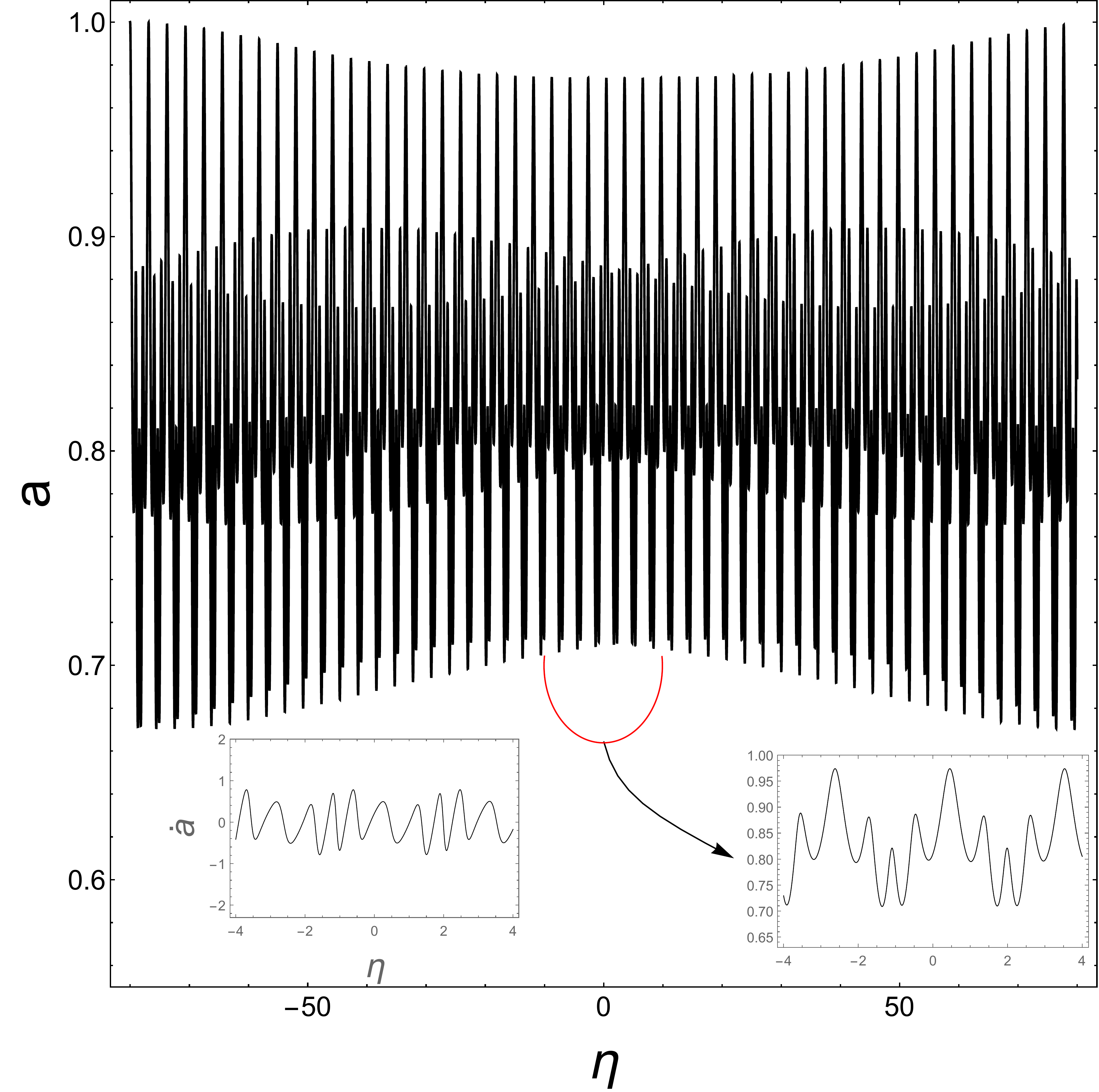}
		\includegraphics[scale=0.2]{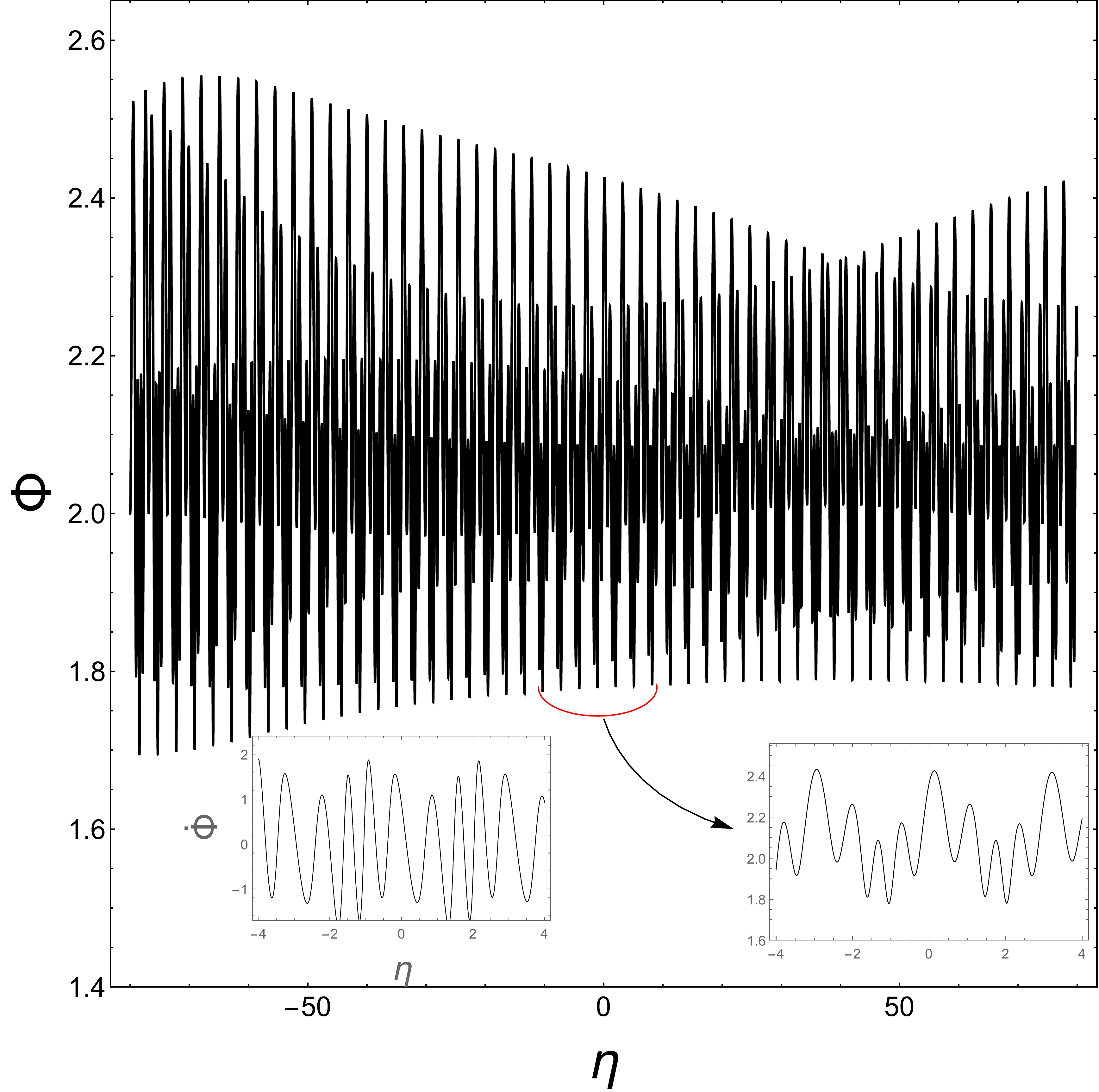}
		\caption{Dynamical evolution of scale factor (left panel) and the {\sf BD} scalar field (right panel) for $\gamma=4/3$, $\omega=-2.6$ and $r=-5$. The initial values for dynamical variables has been set as ${\sf X}_0={\sf Y}_0=0$, ${\sf Z}_0=0.33$ and ${\sf Q}_0=-0.3166$. The right insets show the behavior of $a$ and $\Phi$ within a shorter time interval and the left ones are plotted for time derivatives of these quantities.}\label{figaphi43}
	\end{center}
\end{figure}

\begin{figure}
	\begin{center}
		\includegraphics[scale=0.21]{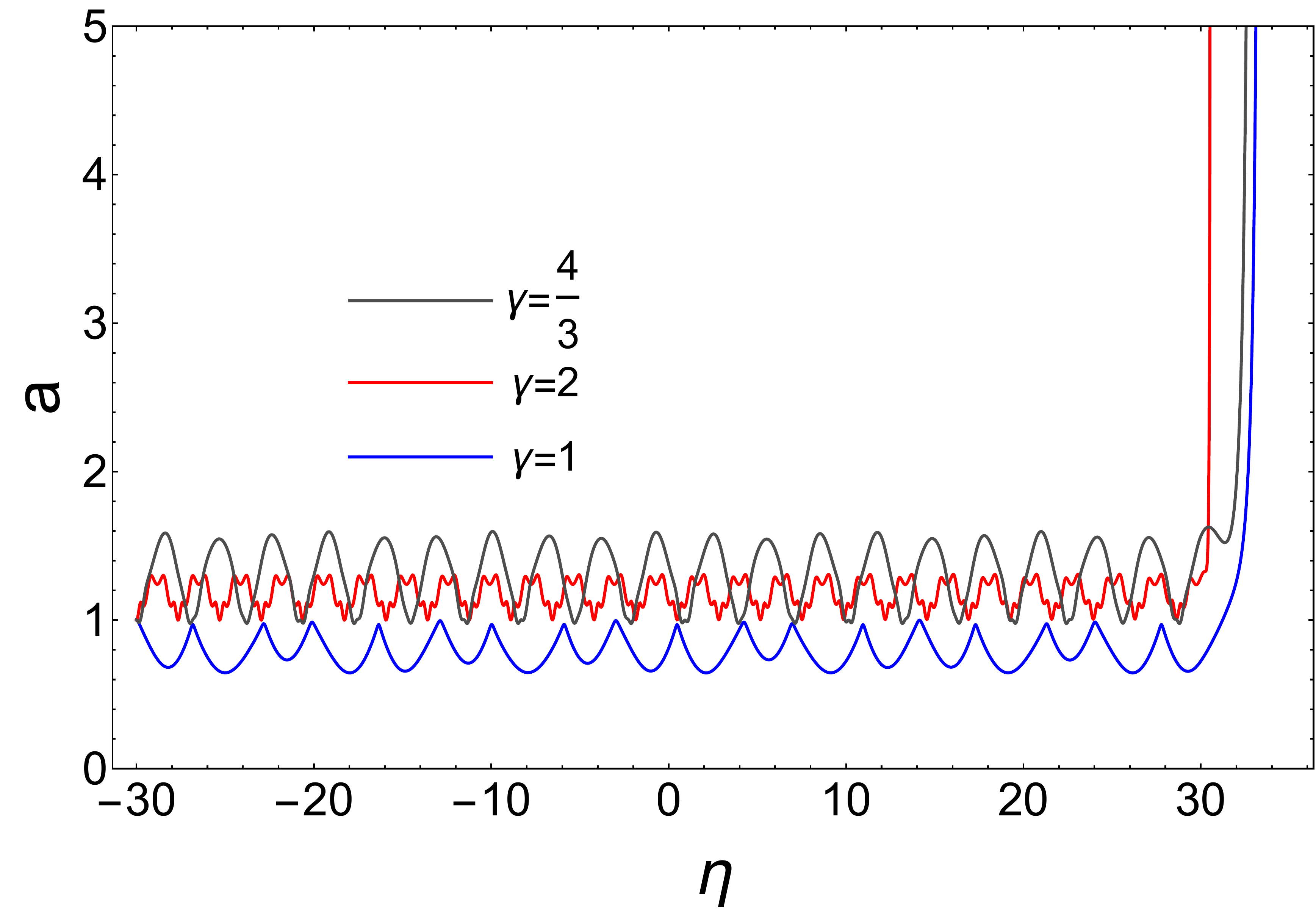}
		\includegraphics[scale=0.21]{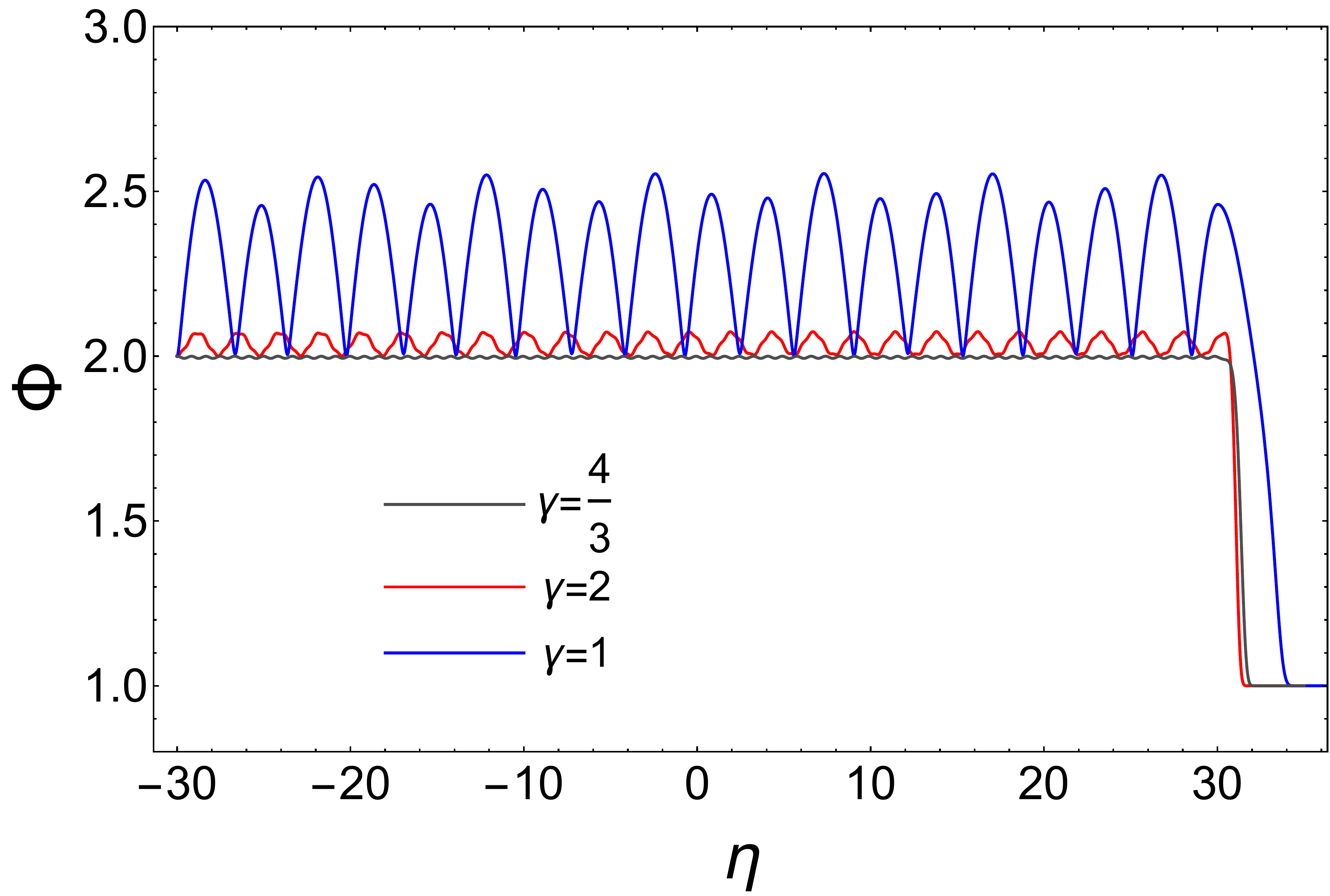}
		\caption{Dynamical evolution of scale factor (left panel) and the {\sf BD} scalar field (right panel) for the same model parameters as specified in Figs. (\ref{fig2})-(\ref{fig4}) but slightly different initial values. We have set the parameters of the expression for {\sf EoS} departure as $\gamma_1=0.001$, $\alpha=1$ and $\eta_0=30000$.}\label{fig5}
	\end{center}
\end{figure}

\begin{figure}
	\begin{center}
		\includegraphics[scale=0.4]{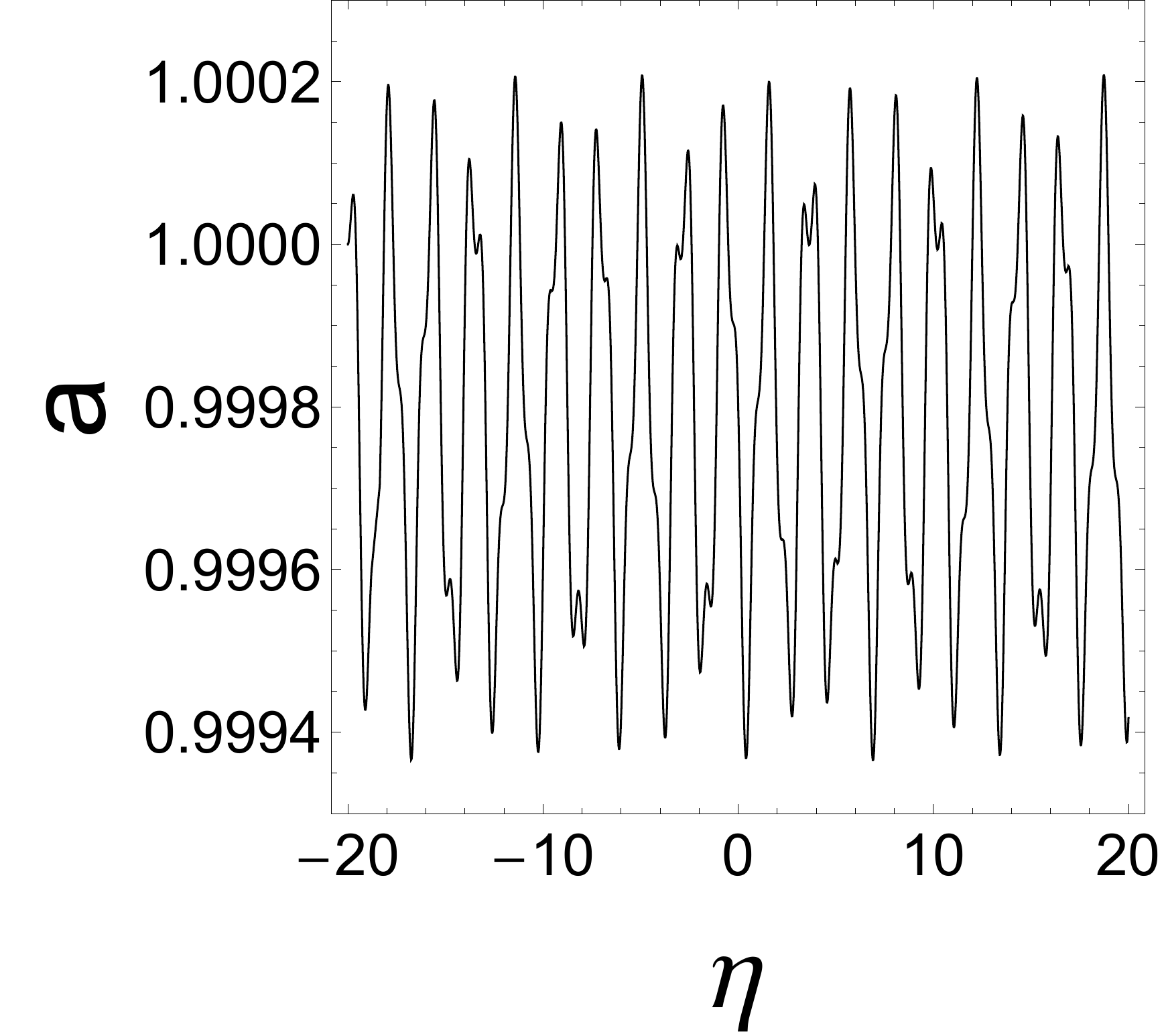}
		\includegraphics[scale=0.4]{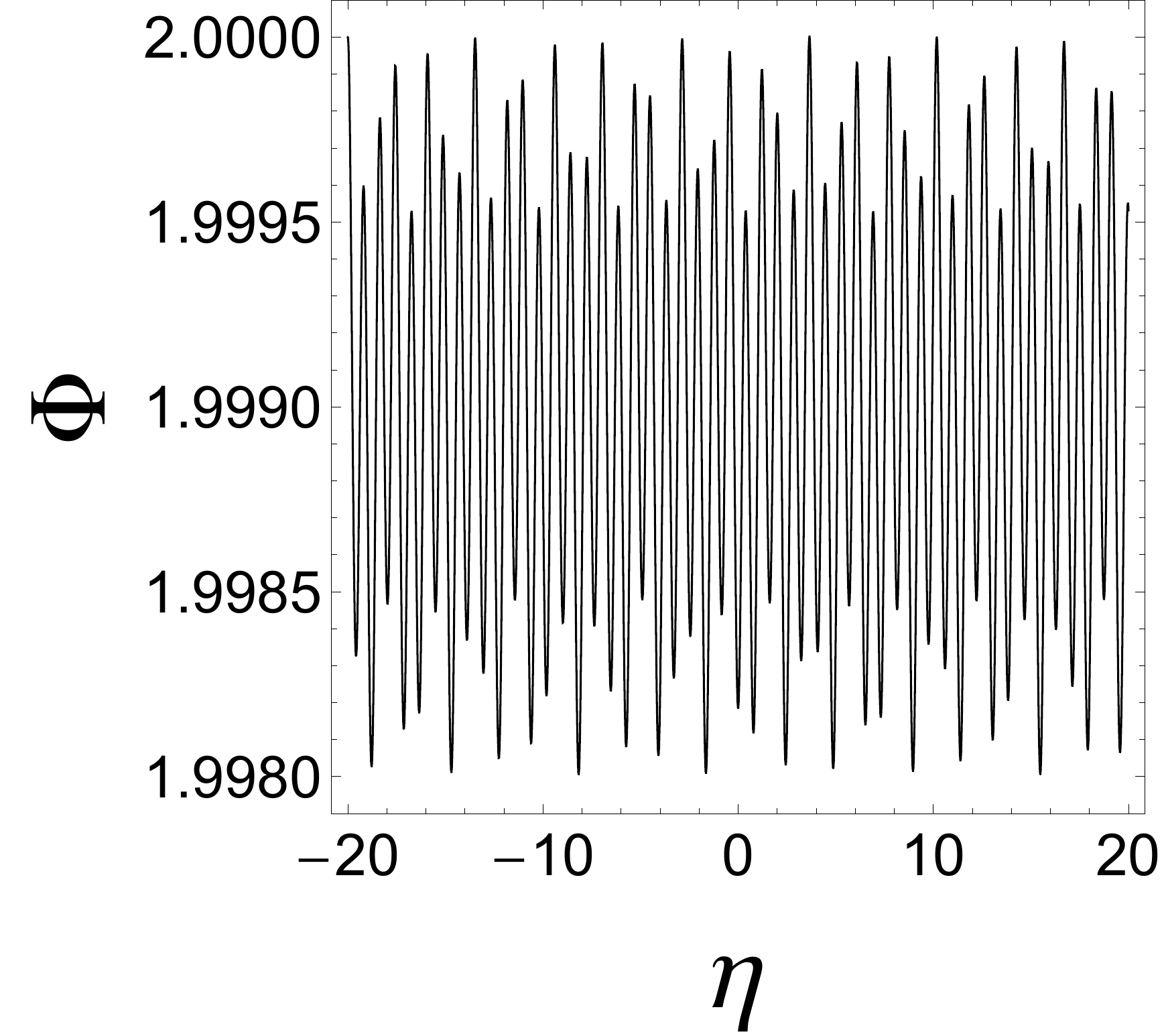}
		\caption{Dynamical evolution of the scale factor (left panel) and the {\sf BD} scalar field (right panel) for
         $\gamma=2$, $\omega=1$ and $r=-9.5$. The initial values for numerical integration has been set as $\sf{X_0}=\sf{Y_0}=0, \sf{Z_0}=2.25, \sf{Q_0}=-0.66$.}\label{figrev}
	\end{center}
\end{figure}

 
As it is not possible to visualize the full 4D phase-space constructed by $({\sf X,Y,Z,Q})$ variables, we proceed to pursue the dynamical evolution of the system in 3D and 2D subspaces of the full phase-space. Figure~(\ref{fig2}) presents 2D and 3D slices of full phase-space for a Universe dominated by a dust fluid. We observe that, given the initial data set, the Universe experiences fluctuations around the center equilibrium point. The 3D simulation gives us better visualizing of the dynamical evolution of the Universe within the subspace of 4D phase-space. We observe that though the amplitude of departures from stability grows for a limited time interval in some regions of the 3D phase-space, the system comes back to the neighborhood of the center fixed point as the time passes and it keeps behaving in this way for later times. In Figs. (\ref{fig3}) and (\ref{fig4}) we have plotted numerical integration of the system (\ref{dyn1})-(\ref{dyn4}) for other {\sf EoS} parameters, where we observe that the overall behavior of the system is same as the dust case but with different orbits. In figures (\ref{figaphi1}), (\ref{figaphi2}) and (\ref{figaphi43}) we have sketched the dynamical evolution of the scale factor and {\sf BD} scalar field where it is seen that these quantities undergo oscillations around their equilibrium values. The scale factor remains finite and nonzero and also the change from an expanding regime to a contracting one occurs smoothly, see the inset diagrams in Figs. (\ref{figaphi1})-(\ref{figaphi43}). This behavior of the scale factor signals that the Kretschmann invariant $({\sf K}=12[(\ddot{a}/a)^2+(\dot{a}/a)^4])$ behaves regularly during the dynamical evolution of the Universe, hence, the {\sf ES} state lives within a nonsingular stable regime, past-eternally. One may also argue that the perturbation modes never damp out, nor do they grow up in time but instead, they continue to exist (with  oscillating and non-growing amplitude) as the Universe evolves. If the perturbation modes go to zero as the time passes, the Universe may undergo an unstoppable collapse process and thus a spacetime singularity may be the end-state of the Universe. In case these modes grow up, inevitable expansion of the Universe could occur so that the {\sf BD} scalar field can go to a vanishing value which would correspond to an infinitely strong gravitational coupling. Hence, we require that the perturbations remain and evolve within the system in order that they could act as a balancing effect and possibly prevent the Universe from instantaneous collapse or expansion, at least, as long as the Universe stays in an static state. However, the Universe have to eventually leave the static state and enters to an inflationary phase. A suitable mechanism which could provide a setting in order that the Universe departs from the static regime is a slight change within the {\sf EoS} parameter so that this parameter turns out to be time dependent temporarily and under this change within the {\sf EoS} parameter, static equilibrium can be broken and the Universe could have chance to escape the static state and eventually enters an inflationary era. The study of how perturbations could affect such a transition may not be an easy task but one could intuitively imagine that the perturbations which have had dynamical evolution along with the Universe could possibly assist it to escape the static regime and begins its inflationary expansion. In order to have an exit to inflation scenario, we assume that the {\sf EoS} parameter turns out to be time dependent with functionality $\gamma(\eta)=\gamma_0+\gamma_1(1-{\rm exp}(\alpha \eta/\eta_0))$ and then solve the system of differential equations (\ref{dyn1})-(\ref{dyn4}) numerically. The results are shown in Fig.~(\ref{fig5}) where we see that after oscillations around its static value, the scale factor begins to increase (with no return) in a short time period allowing thus the Universe to enter an inflationary regime. Meanwhile, having oscillations around its static value, the {\sf BD} scalar field also starts to decrease and finally settles down to a non-vanishing constant value and this value can be set, using a suitable system of units, to the gravitational coupling constant. It is worth mentioning that since in addition to the spin effects, the {\sf BD} scalar field could also act as a source of torsion field, then the {\sf ECBD} theory is reduced to the {\sf EC} theory after the {\sf BD} scalar field comes to a rest at a constant value but the torsion field may not be totally vanished as the spin effects are still present. 

		

\section{Concluding Remarks}\label{summconc}
The study of early Universe physics has been a hot topic in the fields of cosmology and astronomy. Over the past decades, a huge amount of research works have been done in this arena, which have extended our knowledge of the origin and evolution of the Universe. One of the most beautiful and popular cosmological models describing a non-singular state of the early Universe is the {\sf ES} model that the study of which has been extensively carried out in the literature. In the present work we investigated the existence and stability of the {\sf ES} Universe in the context of {\sf ECBD} theory and showed that the corresponding {\sf ES} solution is stable in the sense of dynamically corresponding to a center equilibrium point. Moreover, The solution is stable against small homogeneous perturbations in the sense that these perturbations prompt the scale factor and {\sf BD} scalar field to oscillate about their static values so that the Universe undergoes small departures from its stable phase in the from of contractions and expansions. Since the effects of spin contribution to the field equation (\ref{gmunu}) show themselves as negative pressure, one may argue that the spacetime torsion generated by spin contribution, induces gravitational repulsion in fermionic matter at extremely high densities and tends to destabilize the {\sf ES} state via this repulsive effect and finally breaks down the stability. Examples of cosmological as well as astrophysical nonsingular scenarios have shown that the initial singularity of the Universe or the singularity as the end-state of a gravitational collapse process can be remedied as a result of the repulsive effects due to spin~\cite{torgrabounce}. However, we observed that even if this repulsive effect is considered within the {\sf BD} theory the static stable state for the Universe could exist but with different conditions on model parameters in comparison to the {\sf BD} theory without torsion and spin effects~\cite{emergentBDT}. The conditions on stability as discussed in the present model put some restrictions on the values of {\sf EoS} and {\sf BD} parameters along with a dimensionless parameter which is related to the scalar field potential and its derivative. Hence, it is easy to figure out that the nature of the fixed point crucially depends on model parameters 
so that this dependence could be helpful for providing an exit to inflation scenario, in which, assuming a slowly varying {\sf EoS} parameter for a short time interval, the Universe that has been living in a stable past-eternal static state (a center equilibrium point) could eventually enter into a phase where the stability of the solution is broken leading to an inflationary era. It is not however far fetched to ideate the possibility of having a time dependent {\sf EoS} parameter for a short period of time due to disturbances that the Universe were experiencing. Another possibility is that, as the {\sf BD} coupling parameter bears the ratio of the scalar to tensor couplings to matter in such a way that the larger the values of this parameter the smaller the scalar field effects, one can consider a running {\sf BD} coupling parameter in the sense that it gets smaller values, therefore significant contribution of the scalar field at the early stages of the
Universe, while evolving to larger values at present epoch~\cite{Will-Faranoni}.
\section{Acknowledgments}
This work has been supported financially by Research Institute for Astronomy \& Astrophysics of Maragha (RIAAM) under research project
No. 1/6025-70.

\end{document}